\documentclass[preprintnumbers,superscriptaddress,12pt,tightenlines,nofootinbib]{revtex4}
\usepackage{amsmath}
\usepackage{amssymb}
\usepackage{graphicx}
\usepackage{color}
\usepackage{epsfig}% Include figure files
\usepackage{dcolumn}% Align table columns on decimal point
\usepackage{float}
\usepackage{lipsum}
\usepackage{soul}
\usepackage{bm}
\usepackage{times}
\usepackage[paperwidth=210mm,paperheight=297mm,centering,hmargin=2cm,vmargin=2cm]{geometry}
\usepackage{hyperref}
\hypersetup{
  colorlinks=true,
  citecolor=blue,
  linkcolor=blue,
  urlcolor=blue}

\usepackage{epsfig}% Include figure files
\usepackage{dcolumn}% Align table columns on decimal point
\usepackage{morefloats}
\def\be{\begin{equation}}
\def\ee{\end{equation}}
\def\bea{\begin{eqnarray}}
\def\eea{\end{eqnarray}}
\def\gsim{\ \rlap{\raise 2pt\hbox{$>$}}{\lower 2pt \hbox{$\sim$}}\ }
\def\lsim{\ \rlap{\raise 2pt\hbox{$<$}}{\lower 2pt \hbox{$\sim$}}\ }
\def\dslash{\kern-4pt \not{\hbox{\kern-2pt $\partial$}}}
\def\pslash{\not{\hbox{\kern-2pt p}}}

\def\th23{{$\theta_{23}$}}

\newcommand{\ms}{\Delta m^2_{21}}
\newcommand{\ma}{\Delta m^2_{31}}
\newcommand{\dcp}{\delta_{CP}}
\newcommand{\nova}{NO$\nu$A\ }

\begin{document}

\renewcommand{\arraystretch}{2}
% \ifpdf
%\DeclareGraphicsExtensions{.pdf,.jpg,.mps,.png}
% \else
\DeclareGraphicsExtensions{.eps,.ps}
% \fi

%\preprint{ROME1-1364-2003}

\title{Spotlighting the sensitivities of T2HK,T2HKK and DUNE}
%\title{The role of T2HK and T2HKK and DUNE in determination of  mass hierarchy, octant
%and $\dcp$ }


%%%%%%%%%%%%%%%%%%%%%%%%%%%%%%%%%%%%%%%%%%%%%%%%%%%%%

\author{Kaustav Chakraborty}
\email[Email Address: ]{kaustav@prl.res.in}
\affiliation{Theoretical Physics Division, Physical Research Laboratory, Ahmedabad - 380009, India}
\affiliation{Discipline of Physics, Indian Institute of Technology, Gandhinagar - 382355, India}

\author{K. N. Deepthi}
\email[Email Address: ]{deepthi@prl.res.in}
\affiliation{Theoretical Physics Division, 
Physical Research Laboratory, Ahmedabad - 380009, India}
 
\author{Srubabati Goswami}
\email[Email Address: ]{sruba@prl.res.in}
\affiliation{Theoretical Physics Division, 
Physical Research Laboratory, Ahmedabad - 380009, India}

%\collaboration can be followed by \email, \homepage, 
%\thanks as well.
%\date{\today}
%%%%%%%%%%%%%%%%%%%% abstract %%%%%%%%%%%%%%%%%%%%%%%%%%%%%%%%%%%%%%%%%%
\begin{abstract}
Neutrino oscillation physics has entered the precision era and the potential 
forthcoming experiments Hyper-Kamiokande and Deep Under-ground Neutrino Experiment (DUNE) are expected to lead this endeavor. 
%DUNE has been proposed to run for 10 years with equal neutrino and antineutrino runs. The Hyper-Kamiokande collaboration has recently proposed a new experiment T2HKK as an alternative to T2HK experiment (Tokai to Hyper-Kamiokande). Both T2HK \& T2HKK experiments are proposed to run with a neutrino to antineutrino run time ratio of $1:3$ during a period of 10 years. 
In this paper we perform a comprehensive study of the
octant, mass hierarchy and CP discovery sensitivities of DUNE, T2HK \& T2HKK
in their individual capacity and investigate the synergies of the 
aforementioned experiments with the on going T2K and NO$\nu$A experiments. 
%Furthermore, we present an elaborate discussion on the capability of T2HKK to resolve 
%the octant of $\theta_{23}$.
We present a comparative account of the probabilities at the three baselines 
and explore in detail the physics issues which can cause the discrepancies in 
the sensitivities among the different experiments. 
We also  find out the optimal exposure required by these experiments
for achieving $5\sigma$ hierarchy and octant sensitivity 
and to discover CP violation at
$3\sigma$ for 60\% values of $\dcp$. 
In addition we vary the neutrino-antineutrino runtime 
ratios for T2HK \& T2HKK and check if the sensitivities
are affected significantly due to this.

%In addtion we vary the neutrino-antineutrino 
%proportion for T2HK and T2HKK  and check if other   best neutrino to antineutrino run time ratio for T2HK \& T2HKK in order to maximize the sensitivity of these two experiments to determine neutrino mass hierarchy, octant of $\theta_{23}$ and the Dirac CP phase $\delta_{CP}$.  
%We also compare the corresponding sensitivities of the optimized T2HK and T2HKK experiments with T2K, NOvA and DUNE while assuming their proposed run times.    
%Secondly, we evaluated the combined discovery potential of currently running T2K, NO$\nu$A and the upcoming T2HK, T2HKK and DUNE experiments.
\end{abstract}

\maketitle

\section{Introduction}
Neutrino oscillation physics is an emerging field of research that has been
posing many challenges over the past few decades. Nevertheless, with the help 
of many phenomenal experiments much progress has been made in precisely 
determining the oscillation parameters $\theta_{12}$, $\theta_{13}$, $|\Delta m^2_{31}|$ and $ \Delta m^2_{21}$. This leaves determination of neutrino mass hierarchy i.e. the sign of $\Delta m^2_{31}$, CP phase $\delta_{CP}$ and the octant of $\theta_{23}$ as the primary objectives of the on-going and the upcoming potential neutrino oscillation experiments. 
In a three flavour framework there can be two possible ordering
of the mass eigenstates $m_i$. If $m_1 < m_2 < m_3$ then
we get the normal hierarchy (NH) and if $m_3 < m_2 \approx m_1$ then
it is called the inverted hierarchy (IH).
Octant of $\theta_{23}$ refers to whether $\theta_{23} < 45^\circ$
i.e. it lies in the lower octant (LO)
or if $\theta_{23} > 45^\circ$ i.e it is located in the higher octant (HO).
For $\dcp$ the values 0 and $\pm 180^\circ$ correspond to CP conservation
and $\pm 90^\circ$ corresponds to maximal CP violation. 

In this regard the currently running NO$\nu$A and T2K experiments have recently published some interesting results. 
T2K data shows a preference for $\delta_{CP} = -90^\circ$, 
maximal $\theta_{23}$ and a mild indication for 
normal mass hierarchy \cite{Abe:2017uxa}. Whereas, the combined $\nu_{\mu} \rightarrow \nu_e$ appearance and disappearance channel data of NO$\nu$A \cite{Adamson:2017gxd} has recently 
suggested that for all values of $\delta_{CP}$ inverted mass hierarchy along with $\theta_{23}<45^\circ$ is disfavored at 93$\%$ C.L. and there are two degenerate best-fit points (1) $\sin2\theta_{23} = 0.404$, $\delta_{CP}=266.4^\circ$ (2) $\sin2\theta_{23} = 0.623$, $\delta_{CP}= 133.2^\circ$ when neutrino masses obey normal hierarchy. 
Global analysis of oscillation data indicates $\dcp = -90^\circ$ \cite{Capozzi:2016rtj,Esteban:2016qun,deSalas:2017kay}. 
There is a preference towards HO and a weak hint for NH.
The difficulties in accurate determination of these parameters faced by 
the current experiments 
are the parameter degeneracies allowing for multiple solutions.
%The parameter degeneracies faced by the present experiments 
This can be surpassed
by the future experimental proposals like T2HK \cite{Abe:2014oxa} , T2HKK \cite{Abe:2016ero} and DUNE \cite{Acciarri:2015uup} thereby improving the hierarchy, octant 
and $\dcp$ sensitivity.  
Among these DUNE is a proposed Fermilab based experiment with 
a Liquid Argon Time Projection Chamber 
detector placed at the Sanford Undergrounds Research 
Facility (SURF) which is 1300 km downstream to the initial neutrino beam.
%with a baseline of 
%1300 km, with a Liquid Argon Time Projection Chamber 
%detector placed at the Sanford Undergrounds Research 
%Facility (SURF). 
T2HK is an experiment based in Japan which plans to send neutrino 
beam from Tokai to Kamioka through a baseline of 295 km  
to two Hyper-Kamiokande 
detectors.   
Recently an alternative proposal in which one of the 
detectors in Japan will be shifted to Korea has been mooted. This is termed 
as T2HKK \cite{Abe:2016ero}.   
There have been several studies on the capabilities of 
DUNE \cite{Agarwalla:2013hma,Ghosh:2014rna,
Bora:2014zwa,Barger:2013rha,Deepthi:2014iya,Nath:2015kjg} and 
T2HK \cite{Abe:2014oxa,C.:2014ika,Coloma:2012ji}
for determination of the three major unknowns mentioned above.
The T2HKK proposal studied the hierarchy and $\dcp$ sensitivity
of the set up with respect to three off-axis angles
 $1.5^\circ, 2^\circ, 2.5^\circ$ \cite{Abe:2016ero}.  
With the inception of this proposal,  
the physics possibilities 
of T2HKK regarding determination of hierarchy, octant and $\dcp$ 
has been studied in ref. \cite{Ballett:2016daj, Agarwalla:2017nld, 
Ghosh:2017ged,Raut:2017dbh}.
In ref. \cite{Ballett:2016daj},
the mass hierarchy and CP sensitivity of 
DUNE, T2HK, DUNE+T2HK were studied in detail. 
The also elaborated on  the optimization of alternative designs for DUNE and 
the T2HKK proposal. 
In \cite{Agarwalla:2017nld} a hybrid setup in which the antineutrino run 
of T2HK is substituted by antineutrinos coming from muon decay at rest ($\mu$-DAR) has been studied for determination of hierarchy, octant and $\dcp$.  
%the mass hierarchy sensitivity, CP violation sensitivity, precision measurements of $\delta_{CP}$, octant degeneracy and the precision of $\theta_{23}$
%by taking into account their recently updated proposals. 
The author of ref. \cite{Raut:2017dbh} has studied the sensitivity of T2HK, 
T2HKK and DUNE to determine mass hierarchy and to measure the CP phase 
$\delta_{CP}$ 
and also made a comparative analysis of DUNE+T2HK and DUNE+T2HKK. 
%However, octant sensitivity of T2HKK was not studied in ref. \cite{Raut:2017dbh}. 
In ref. \cite{Ghosh:2017ged} the role of systematic uncertainties in 
the determination of  hierarchy, octant and $\dcp$ in the three set ups 
were studied. However octant sensitivity of T2HKK 
and a detailed comparative study 
of the three setups have not been presented in any of these references.

In this work, one of our major aims is to study the octant sensitivity 
of the T2HKK setup. It is well known that one of the challenges 
for precise determination of $\dcp$ is the octant-$\dcp$ degeneracy.
Therefore the CP discovery potential of a set up is intimately 
connected with its octant resolution capability if such degeneracies 
are present. Thus it is important to investigate to what extent 
an experiment can resolve these degeneracies. 
With this aim we do a comprehensive analysis of the octant 
discovery potential of T2HKK and compare it with the same for 
the other two setups. Further, we give a detailed account of the 
underlying physics reasons which are causing differences 
in the octant sensitivity of these three experiments. 
Additionally we also do a comparative analysis of the hierarchy and 
CP discovery potential of the three set ups  
for the sake of completeness.
%Although the hierarchy and CP sensitivity of all the three set ups have 
%been discussed elsewhere there is no single work which have 
%studied all the three experiments together. 
Thus our work  provides a ready 
reference for the comparison amongst the hierarchy, CP discovery and octant 
sensitivities and the underlying physics issues of these facilities. 
In view of the fact that by the time these experiments are 
operative, the current experiments T2K and \nova will 
already have their results 
we also show to what extent the inclusion of this information can improve 
the sensitivities.
In addition, we also estimate the optimal 
exposures/run times of these setups to achieve $5\sigma$ sensitivity in 
octant and hierarchy determination and $3\sigma$ sensitivity to CP discovery 
potential for 60\%  values of the CP phase $\dcp$. 
Furthermore we consider two different neutrino and antineutrino
runtime ratios (3:1 and 1:1)  in T2HK and T2HKK apart from the 
proposed 1:3 ratio and study to what extent the sensitivities are 
affected. 
%of $\nu$ and $\bar{\nu}$ in T2HK and T2HKK experiment give the 
%best results   other combinations of runtime ratios.    

%%%%%%%%%%%%%%%%%%%%%%%%%%%%%%%%%%%%%%%%%%%%

The outline of the paper is as follows. In  section 
\ref{prob-discuss} we present the relevant probabilities 
%and compare the hierarchy, octant, 
%CP sensitivities of these facilities in view of the  probabilities. 
and discuss the degeneracies that are faced by the
different experiments in terms of probabilities. 
The following section contains the results on octant, hierarchy and CP 
discovery potentials of the three experiments with and without inclusion
of T2K and \nova results. A concise discussion regarding the experimental
setups have also been included in this section. 
In section \ref{optimal} we obtain the optimal exposures required by 
various setups to achieve $5\sigma$ octant and hierarchy sensitivity. 
We also present the exposure needed to achieve $3\sigma$ sensitivity to 
discovery of CP violation for 60\% values of $\dcp$. 
In the next section we study the effect of varying the proportion of 
neutrino and antineutrino run times in the T2HK and T2HKK experiment 
to explore if this can cause any discernible effect as opposed to
the proposed runtime ratio. 
We make our concluding remarks in the final section.

%%%%%%%%%%%%%%%%%%%%%%%%%%%% Table %%%%%%%%%%%%%%%%%%%%%%%%%%%%%%%%%%
\begin{table}[h!]
\begin{center}
\begin{tabular}{|c|c|c|}
\hline
Oscillation parameters & True value & Test value \\
%\\
\hline
$ \sin^{2}2 \theta_{13} $ & 0.085 & 0.07 -- 0.1 \\
$ \sin^{2}\theta_{12} $ & 0.304 & -- \\
$ \theta_{23} $ & $42^\circ$(LO), $48^\circ$(HO) & $39^\circ-51^\circ$  \\
$\Delta m^2_{21} $ & 7.40 $ \times 10^{-5} eV^{2} $ & -- \\
$\Delta m^2_{31} $ & 2.50 $ \times 10^{-3}eV^{2}  $ & (2.35 -- 2.65) $ \times 10^{-3} eV^{2} $ \\
$ \delta_{CP} $ & $ -180^\circ $ to $ +180^\circ $  & $ -180^\circ $ to $ +180^\circ $ \\
\hline
\end{tabular}
%\vspace{3mm}
\caption{True values and marginalization ranges of neutrino oscillation parameters used in our
numerical analysis.}
\label{param_values}
%\caption{True values and marginalization ranges of neutrino oscillation parameters used in our 
%numerical analysis.}
%\label{param_values}
\end{center}
\end{table}

%%%%%%%%%%%%%%%%%%%%%%%%%%%%%%%%%%%%%%%%%%%%%%%%%%%%%%%%%%%%
\section{Probability discussions} \label{prob-discuss}
%%%%%%%%%%%%%%%%%%%%%%%%%%%%%%%%%%%%%%%%%%%%%%%%%%%%%%%%%%%%%%%

%%%%%%%%%%%%%%%%%%%%%%%%%%%%%%%%%%%%%%%%%%%%%%%%%%%%%%%%%%%%%%%%%%%%%%%%%%%%%%%
\begin{figure}[h!]
%\vspace{-1.4cm} 
        \begin{tabular}{clr}
                \hspace*{-0.65in} 
                &
                \includegraphics[width=0.5\textwidth]{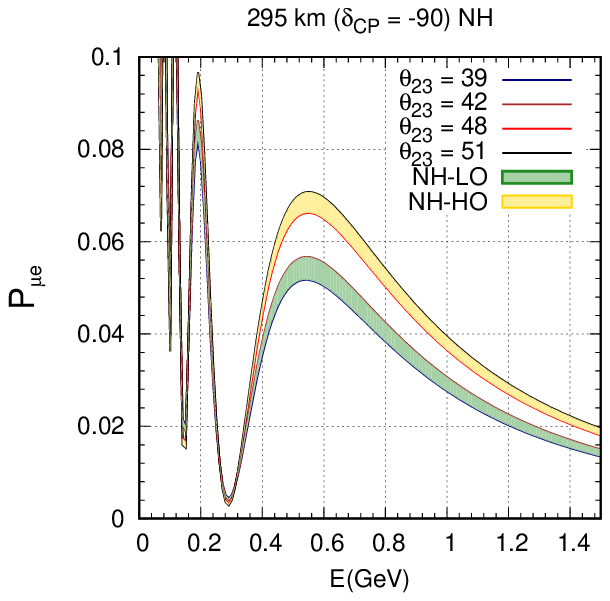}              
                \hspace*{-1.0in}
                \includegraphics[width=0.5\textwidth]{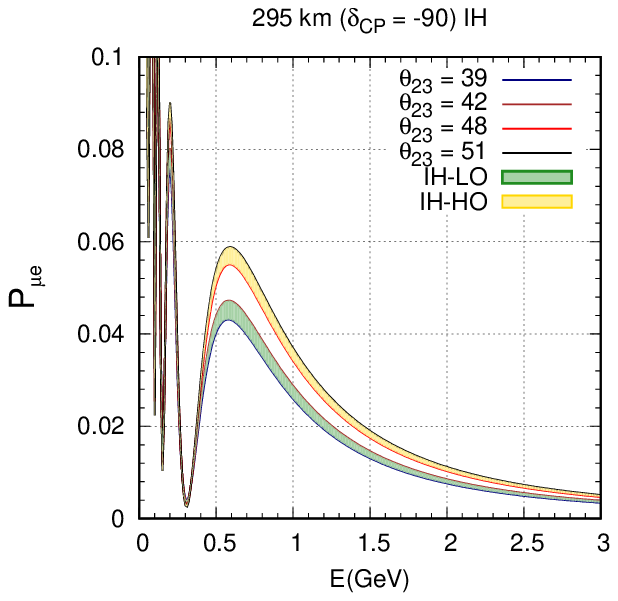}\\
 %               \hspace*{-1.0in} 
                &
                \includegraphics[width=0.5\textwidth]{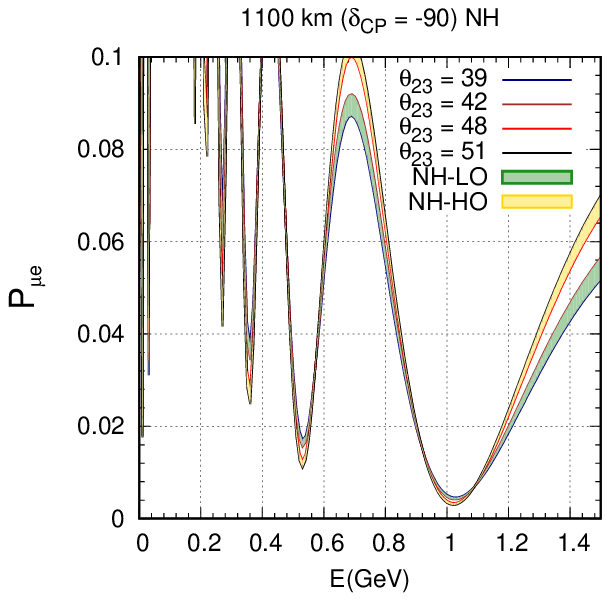}               
                \hspace*{-1.0in} 
                \includegraphics[width=0.5\textwidth]{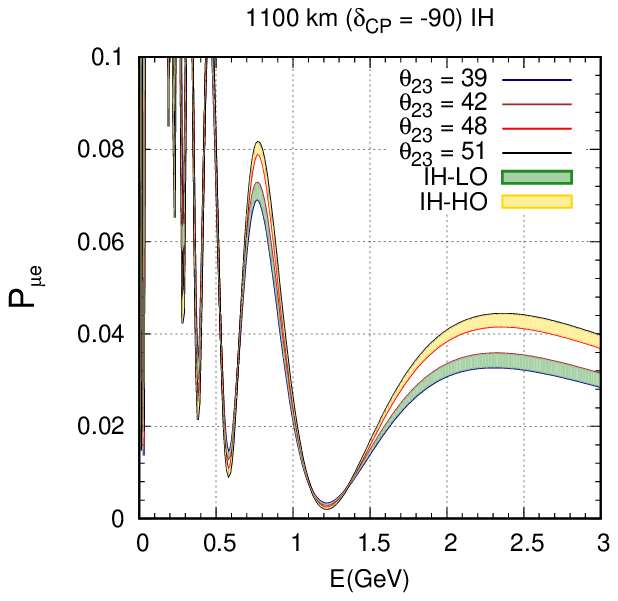}\\
                &
                \includegraphics[width=0.5\textwidth]{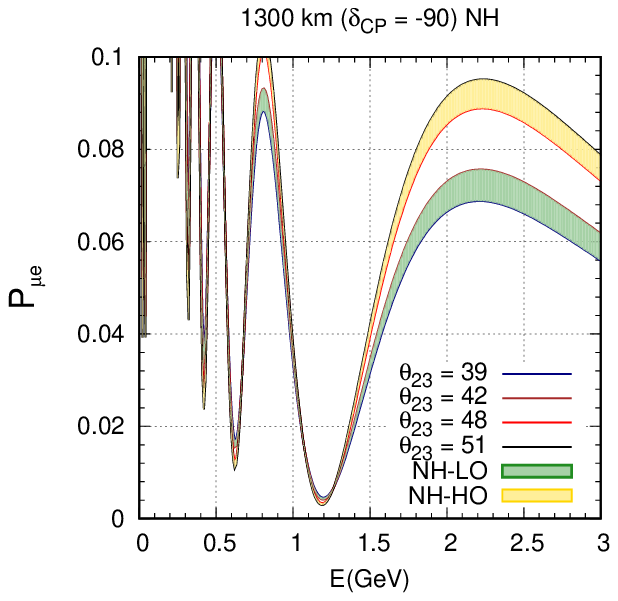}               
                \hspace*{-1.0in} 
                \includegraphics[width=0.5\textwidth]{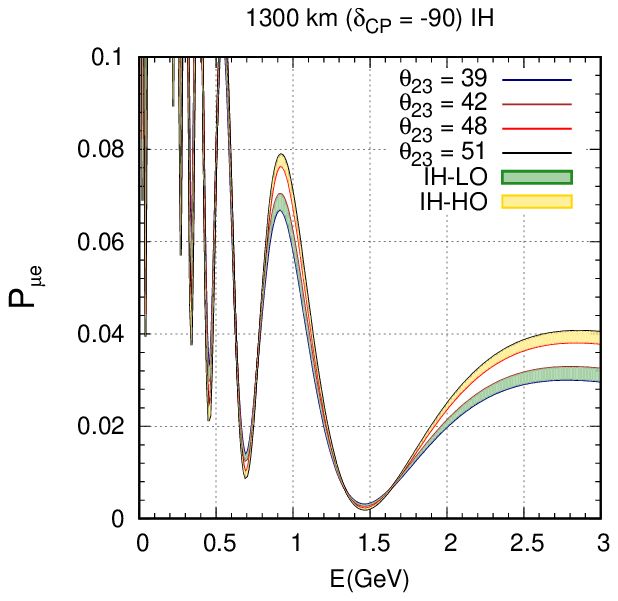}              
                \end{tabular}
\caption{Appearance probabilities $P_{\mu e}$ vs Energy for 295 km, 1100 km and 
1300 km. The left panel is for NH and right panel for IH. The bands 
are due  to variation over $\theta_{23}$. Here $\dcp$ and $\theta_{13}$ are in degrees.  }
\label{prob-vs-E}
\end{figure}
%%%%%%%%%%%%%%%%%%%%%%%%%%%%%%%%%%%%%%%%%%%%%%%%%%%%%%%%%%%%%%%%%%%%%%%%%%%%%%%

%%%%%%%%%%%%%%%%%%%%%%%%%%%%%%%%%%%%%%%%%%%%%%%%%%%%%%%%%%%%%%%%%%%%%%%%%
\begin{figure}[h!]
%\vspace{-1.4cm} 
%        \begin{tabular}{clr}
                \hspace*{-0.65in} 
%                &
                \includegraphics[width=0.5\textwidth]{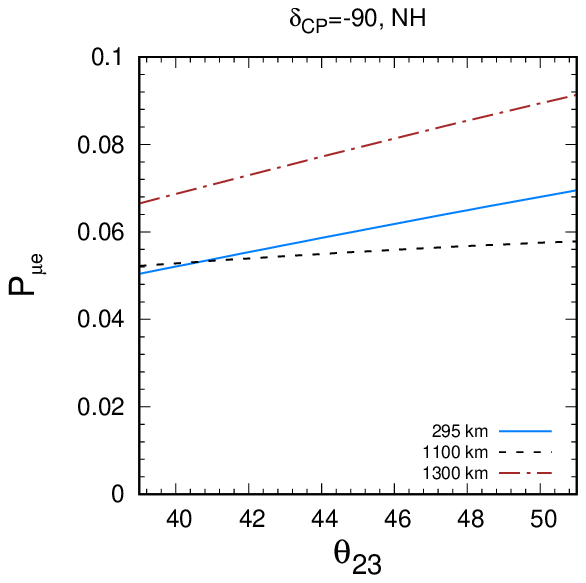}              
                \hspace*{-1.0in}
                \includegraphics[width=0.5\textwidth]{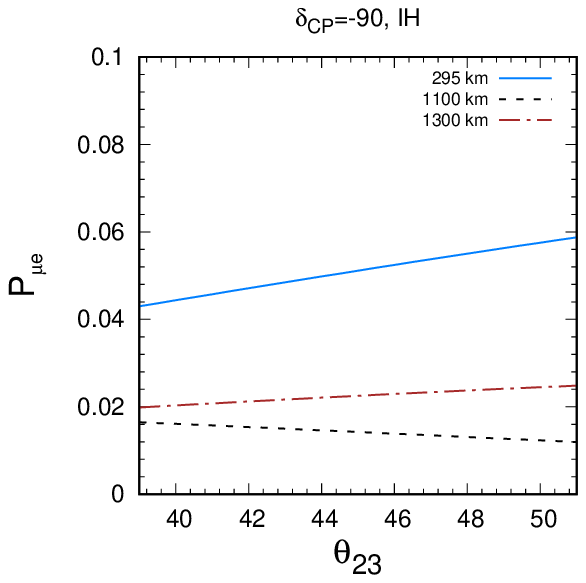}
\caption{Appearance probabilities : $P_{\mu e}$vs $\theta_{23}$ for NH and IH. 
$\dcp$ and $\theta_{13}$ are in degrees.}
\label{prob-vs-th23}
\end{figure}
%%%%%%%%%%%%%%%%%%%%%%%%%%%%%%%%%%%%%%%%%%%%%%%%%%%%%%%%%%%%%%%%%%%%%%%%%%%

\begin{figure}[h!]
%\vspace{-1.4cm} 
        \begin{tabular}{clr}
                \hspace*{-0.65in}
                &
                \includegraphics[width=0.5\textwidth]{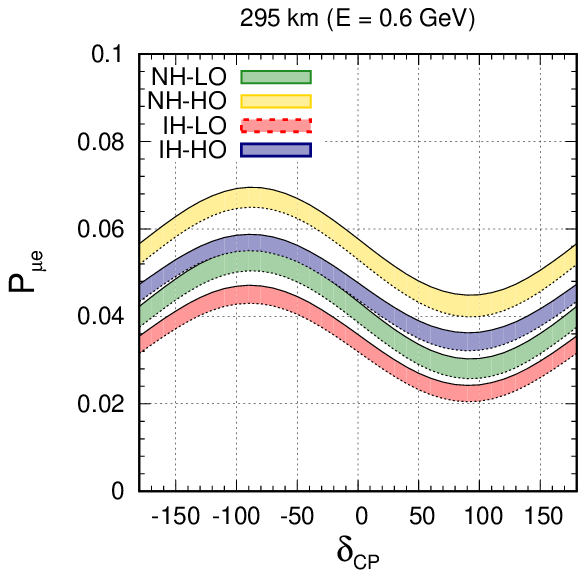}
                \hspace*{-1.0in}
                \includegraphics[width=0.5\textwidth]{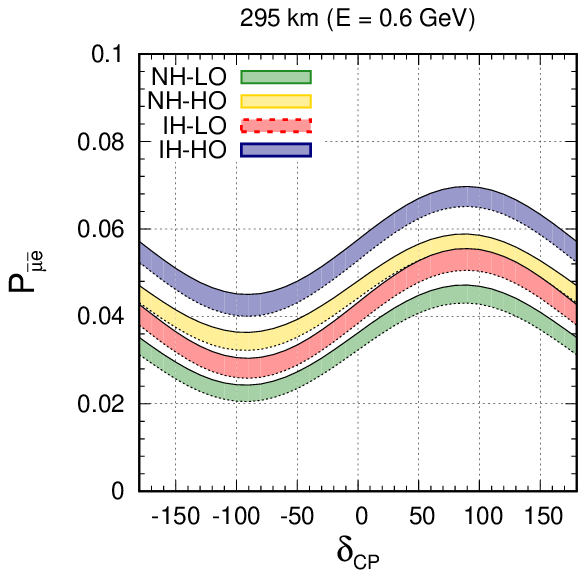}\\
 %               \hspace*{-1.0in} 
                &
                \includegraphics[width=0.5\textwidth]{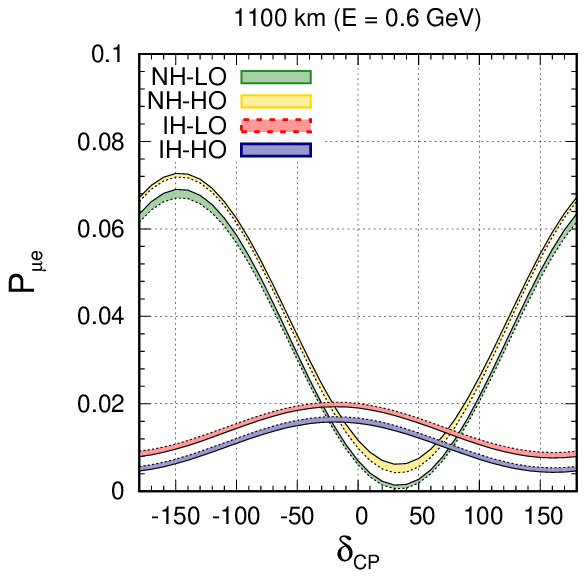}
                \hspace*{-1.0in}
                \includegraphics[width=0.5\textwidth]{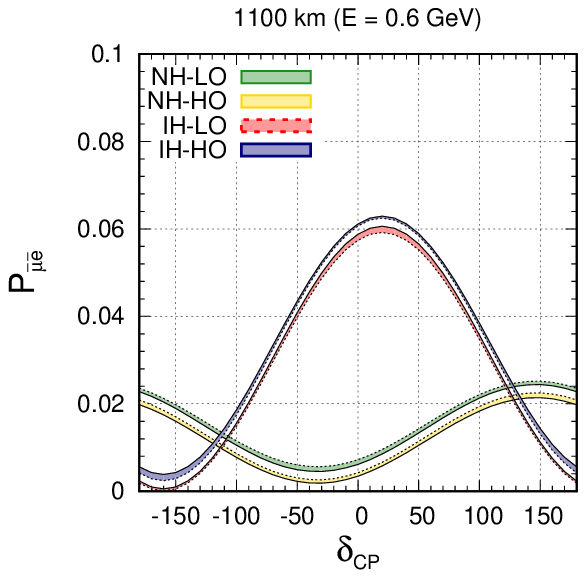}\\
                &
                \includegraphics[width=0.5\textwidth]{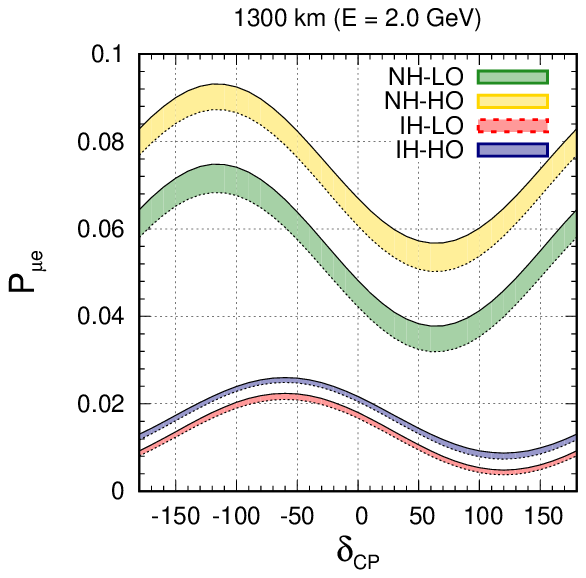}
                \hspace*{-1.0in}
                \includegraphics[width=0.5\textwidth]{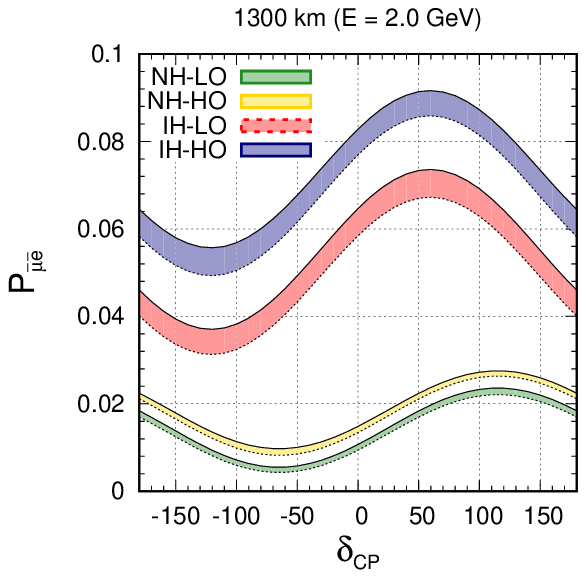}
                \end{tabular}
\caption{Appearance probabilities vs $\dcp$ (degrees) for 295 km,1100 km and 1300 km. 
The left panel is for neutrinos while the right panel is for antineutrinos. 
Each panel contains plots for both NH and IH. 
The band represents variation over $\theta_{23}$.}  
\label{prob-plots}
\end{figure}

%%%%%%%%%%%%%%%%%%%%%%%%%%%%%%%%%%%%%%%%%%%%%%%%%%%%%%%%%%%%%%%%%%%%%%%%%%

Neutrino oscillation experiments measure the 
event rates which in turn depends on the probabilities. 
The probability relevant for determination of the 
three unknowns in the 
long baseline experiments is the 
appearance probability $P_{\mu e}$. 
An approximate analytic expression for this can be obtained  in matter of 
constant density by expanding in terms of two small parameters $\alpha$ and $\sin\theta_{13}$ as \cite{Akhmedov:2004ny}, 
\begin{align}
 P_{\mu e} = ~ & 4 s_{13}^2 s_{23}^2 \frac{\sin^2(\hat{A} - 1) \Delta}{(\hat{A} - 1)^2} \\ \nonumber
           & + 2 \alpha s_{13} \sin2\theta_{12} \sin2\theta_{23} \cos(\Delta + \delta_{CP}) \frac{\sin\hat{A}\Delta}{\hat{A}} \frac{\sin(\hat{A} - 1)\Delta}{\hat{A} - 1} \\ \nonumber
           & + \alpha^2 \sin^22\theta_{12} c_{23}^2 \frac{\sin^2\hat{A}\Delta}{\hat{A}^2},
\end{align}
%{\small \begin{eqnarray}
%\nonumber P\left({\nu_{\mu}(\bar{\nu_{\mu}})}\to \nu_{e}(\bar{\nu_{e}})\right) &=& P_{\mu e}(P_{\bar{\mu} \bar{e}}) = 
%\sin^2 2\theta_{13}\sin^2\theta_{23}
%\frac{\sin^2[(1-(+)\hat{A})\Delta]}{(1-(+)\hat{A})^2} \nonumber \\
%&& 
%+ \, \alpha\cos\theta_{13}\sin2\theta_{12}\sin2\theta_{13}\sin2\theta_{23}\cos\Delta\cos\dcp\frac{\sin(\hat{A}\Delta)}{\hat{A}}
%\frac{\sin[(1-(+)\hat{A})\Delta]}{(1-(+)\hat{A})} \nonumber \\
%&& 
%-(+) \, \alpha\cos\theta_{13}\sin2\theta_{12}\sin2\theta_{13}\sin2\theta_{23}\sin\Delta\sin\dcp\frac{\sin(\hat{A}\Delta)}{\hat{A}}
%\frac{\sin[(1-(+)\hat{A})\Delta]}{(1-(+)\hat{A})} \nonumber \\
%&&
%+ \, \alpha^2\sin^22\theta_{12}\cos^2\theta_{13}\cos^2\theta_{23}\frac{\sin^2(\hat{A}\Delta)}{{\hat{A}}^2} \,.
%\label{eq:pmue}
%\end{eqnarray}
where, $\alpha=\ms/\ma$, $\Delta=\ma L/4E$ and $\hat{A}=A/\ma$. Here
$A=2\sqrt{2} G_F N_e E$ is the matter potential, $N_e$ is the electron number density inside the Earth, $E$ is the neutrino energy and $L$ is the baseline length. For antineutrinos the matter potential changes sign which implies
$\hat{A} \rightarrow -A$ and $\dcp \rightarrow -\dcp$.  
For IH, $\Delta \rightarrow -\Delta$.
%, $\alpha \rightarrow -\alpha$.   

Figure \ref{prob-vs-E} shows the behaviour of the probabilities as a function
of energy for the baselines 295 km, 1100 km and 1300 km. 
These plots are done for a fixed value of $\dcp = -90^\circ$. 
The bands correspond to variation over the octant of $\theta_{23}$ in the range
$39^\circ - 42^\circ$ for LO and $48^\circ - 51^\circ$ for HO.    
The figure shows that for NH the bands are wider for T2HK and DUNE baseline 
as compared to the T2HKK baseline. This indicates that the variation 
of the probability over octant is more for T2HK and DUNE than in T2HKK.
For IH, the bands at the energies where 
the respective flux peaks for T2HK are wider but the bands for T2HKK and DUNE 
are narrower. 

In order to elucidate this further we plot $P_{\mu e}$ as a function of 
$\theta_{23}$ for different baselines of T2K \& T2HK (295 km, 0.6 GeV), 
T2HKK (1100 km, 0.6 GeV ), DUNE (1300 km, 2 GeV) in fig. (\ref{prob-vs-th23}), 
by fixing the corresponding energies at the 
values where the respective flux peaks. 
It is to be noted that 0.6 GeV  corresponds to the first oscillation maxima 
for T2HK baseline while it is close to the second oscillation maxima for the 
T2HKK baseline.
%the energies 0.6 GeV for T2HK \& T2HKK and 
%2 GeV for DUNE.  
The left (right) plot corresponds to $P_{\mu e}$ vs $\theta_{23}$
assuming normal (inverted) hierarchy.
It can be seen that the lines corresponding to the baselines
of 295 km (blue-solid), 1300 km (brown-dash-dotted)
%, 810 km (magenta-dotted) 
have positive
slopes both for NH and IH. However, the curve for 1100 km (black-dashed) is much flatter 
with a small positive slope for NH and negative for IH.

With a view to understand this behaviour of the probability 
with $\theta_{23}$  at different 
baselines we write the probability expression 
eqn. \ref{eq:pmuenew} in the following form 
\cite{Agarwalla:2013ju} 
\begin{eqnarray}
P_{\mu e}=(\beta_{1} - \beta_3) \sin^2\theta_{23} + \beta_{2} \sin 2\theta_{23} \cos(\Delta+\dcp) + \beta_{3}
\label{eq:pmuenew} 
\end{eqnarray} 
where, 
\begin{eqnarray}
\beta_{1} &=& \sin^2 2{\theta_{13}}
\frac{\sin^2\Delta(1 - \hat{A})}{(1 - \hat{A})^2},  \nonumber \\
\beta_{2} &=& \alpha\cos{\theta_{13}}\sin2{\theta_{12}}\sin2{\theta_{13}}\frac{\sin\Delta\hat{A}}{\hat{A}}
\frac{\sin\Delta(1-\hat{A})}{1-\hat{A}}, \nonumber \\
\beta_{3} &=& \alpha^2\sin^22{\theta_{12}}\cos^2{\theta_{13}}\frac{\sin^2\Delta\hat{A}}{\hat{A}^2}
\end{eqnarray}

The $\theta_{23}$ dependence  is seen to come  from the first term of 
eq.(\ref{eq:pmuenew}), varying linearly with $\theta_{23}$ 
with a slope given by ($\beta_1 - \beta_3$).  
Note that over the range of $\theta_{23}$ spanning 
$39^\circ - 51^\circ$, $\sin2\theta_{23}$ stays close to 1 and so
the second term of eq.(\ref{eq:pmuenew})
does not affect the behaviour of $P_{\mu e}$ vs $\theta_{23}$.   
The $\beta_{i}$s for the three different baselines are tabulated in 
table \ref{tab:beta} for both the hierarchies.
%for energies 0.6 GeV 
%(for T2HK and T2HKK) and 2 GeV (for DUNE) which correspond to the 
%energy where the flux peaks. 
The first column corresponds to the L/E ratios of different setups.
For the baselines 295 km and 1100 km the peak energies being the same the 
main difference between the $\beta_i$s enumerated in table \ref{tab:beta} 
is due to the L/E ratio which is much higher for the 1100 km baseline. 
Between the 1300 km baseline and 295 km baseline the L/E ratio is 
approximately the same order and the $\beta_i$s are different 
due to the different energies. 
The higher energy implies a higher value for the matter potential
$\hat{A}$ for the 
1300 km baseline  
entailing $(1 - \hat{A})$ to be smaller and hence $\beta_1$ larger. 
This can be seen from the second column of table \ref{tab:beta}. 
Whereas for 1100 km and 1300 km baselines the energies 
as well as the different $L/E$ ratios 
attribute to the difference in the $\beta_i$ factors. 
From the third column of \ref{tab:beta} 
it can be seen that the $\beta_3$ values of 
1300 km and 295 km are smaller than $\beta_1$ and $\beta_2$.
%it is seen that for the 1300 km and 295 km  
%$\beta_3$ which occurs with the subleading term, 
%is much smaller than the other two terms.  
Thus the probabilities of these two experiments are expected 
to rise as $\sim \sin^2\theta_{23}$ with slope given by $ \approx \beta_1$.
This can be seen from the brown (dash-dotted) and the blue (solid) lines
respectively. 
Comparing these two cases for NH it can be seen that
since $\beta_1$ is higher for 
the 1300 km baseline the variation with $\theta_{23}$ is more than 
that of 295 km. 
%This can be seen from brown (dash-dotted) and the blue (solid)
%lines in the left panel of fig. \ref{prob-vs-th23}.
For IH, on the other hand, $\beta_1$ is much smaller 
for the 1300 km baseline (for the neutrinos) when compared to 295 km baseline
 due to the suppression caused by  matter effect. 
 Hence the variation of $P_{\mu e}$ with $\theta_{23}$ 
is much less for 1300 km when compared to 
295 km as can be seen from
right panel of fig. \ref{prob-vs-th23}.   
%For 295 km also the probability rises with increasing $\theta_{23}$ 
%with slope $ \approx \beta_1$. But matter effect being much less at this 
%distance,  the $\beta_{1}$ for IH is smaller than that of the 1300 km baseline. 
%Thus for IH, the variation with $\theta_{23}$ is more for the 295 km baseline 
%as compared to the 1300 km baseline. This can also be observed in the right  
%panel of fig. \ref{prob-vs-th23}.   
For the 1100 km baseline and NH $\beta_3$ term is 
comparable to $\beta_1$ term 
which makes the $\theta_{23}$ variation much flatter. 
For IH, $\beta_1$ is less than $\beta_3$  and the probability decreases 
with $\theta_{23}$. These patterns can be seen from the black (dashed)
lines in the fig. \ref{prob-vs-th23}.

% it is higher for the 1300 km baseline 
%as compared to the 295 km baseline because of enhanced matter effects 
%for NH. For T2HKK baseline on the other hand because of a different
%L/E value than T2HK the $\Delta$ term is different and $\beta_1$ is less. 
%The $\beta_3$ term on the other hand is  comparable to the $\beta_1$ term
%because of a higher L/E ratio in T2HKK  
%\footnote{Because the $\beta_3$ term is of the same order as the $\beta_1$ 
%term the expression in \ref{eq:pmue} is not very accurate for the T2HKK.}.
%{\bf{need to check this point}} 
%On the other hand for IH $\beta_{1}$ for 1300 km baseline is 
%smaller than that of the 295 km baseline since the denominator 
%$(1+\hat{A})$ is larger for DUNE.  
%For 1100 km baseline this term is much smaller than that of T2HK 
%because of a different L/E.  

\begin{table}[h!]
\begin{center}
\begin{tabular}{|c|c|c|c|c|}
\hline
 & L/E (km/GeV) & $ \beta_1 $ & $\beta_2$  & $\beta_3$  \\
\hline
& $~~~~~$ & NH $~~~~~$ IH & NH $~~~~~$ IH  & NH $~~~~~$ IH  \\
\hline
1300 km & $\sim$ 650  & 0.122 $~~~~~$ 0.028 & 0.018 $~~~~~$ -0.009 & 0.003 $~~~~~$ 0.003 \\
295 km & $\sim$ 490 & 0.094 $~~~~~$ 0.077 & 0.013 $~~~~~$ -0.011 & 0.002 $~~~~~$ 0.002 \\
1100 km & $\sim$ 1800 & 0.045 $~~~~~$ 0.002 & -0.032 $~~~~~$ 0.007 & 0.023 $~~~~~$ 0.023\\
\hline
\end{tabular}
\caption{$ \beta_1 $, $\beta_2$ \& $\beta_3$ values for DUNE, 295km \& 1100km baselines for T2HK \& T2HKK.}
\label{tab:beta}
%\vspace{3mm}
\end{center}
\end{table} 

Figure \ref{prob-plots} shows the probabilities as a function of 
$\dcp$  for the baselines 
295 km, 1100 km and 1300 km for both neutrinos and 
antineutrinos. 
Each figure has four probability bands green, yellow, red and  blue  
corresponding to NH-LO, NH-HO, IH-LO, IH-HO respectively. 
%where NH(IH) signifies Normal(Inverted) Hierarchy \& LO(HO) signifies 
%Lower(Higher) Octant. 
The bands are obtained due to the variation in 
$\theta_{23}$ in the range $35^\circ-42^\circ$ in LO 
and $48^\circ-51^\circ$ in HO. 
This range suffices since in any  case the position of the minima 
is determined by the disappearance channel to be near
 $\approx 90^\circ - \theta_{23}$. 
These kind of plots are helpful in understanding the various degeneracies
at a probability level for values of $\theta_{23}$ not very near to 
maximal mixing. 
An overlap between the bands for different values of $\dcp$ (as can be 
seen by drawing a horizontal line through the bands) would indicate 
a degenerate solution with a wrong $\dcp$ value, whereas, intersection
between the bands would signify a degenerate solution even with the same
value of $\dcp$ \cite{Ghosh:2015ena}. 
For instance from the left hand plot in the first row 
one can see that NH-LO  (green band) is degenerate with IH-HO (blue band) 
in the range
$-180^\circ < \dcp< 0$  (lower-half plane, LHP)  
%whereas the range
% $0<\dcp<180^\circ$ 
%is called the upper-half plane (UHP).}. 
for the 295 km baseline. 
This gives rise to the wrong hierarchy(WH) -- wrong octant (WO) solutions. 
On the other hand NH-HO (yellow band)  in  the LHP 
and IH-LO (red band) in the  
range 
$0<\dcp<180^\circ$ 
(upper-half plane, UHP)  
are devoid of any degeneracy for the neutrinos. 
For the antineutrinos, as can be seen from the right panel of the first row 
NH-LO (green band)  in LHP and IH-HO (blue band)  in UHP are non-degenerate. 
Thus the octant degeneracy is seen to be opposite for neutrinos  
and antineutrinos  and hence combined neutrino and antineutrino run is 
expected to resolve wrong octant solutions. 
Again, considering $\dcp \sim 90^\circ$ in UHP, NH-HO (yellow band) in UHP
is degenerate with IH-HO (blue-band) around $\dcp \sim 0$. This corresponds
to wrong hierarchy-right octant solution. This kind of degeneracy also 
exists for antineutrinos and hence inclusion of antineutrino run cannot 
resolve the wrong hierarchy-right octant solutions. 
Similar conclusions are true for the DUNE baseline
as can be seen from the figures in the bottom row. 
However, because of enhanced matter effects DUNE has much better hierarchy 
sensitivity and therefore the difference between NH and IH bands are much more
for this case  and wrong hierarchy solutions are not seen in the probability 
plots.   
However, for e.g. the wrong octant solutions between 
NH-HO (yellow band) and $\dcp \sim 90^\circ$  and   
NH-LO (green band) and $\dcp \sim 0$ is visible from the neutrino probability 
presented in the left-panel of the bottom row for DUNE 
%and top row for T2HKK 
baseline. But for antineutrinos this degeneracy is not present. 
%So combined neutrino and antineutrino run can give good octant resolution 
%in both experiments. 
 
For the T2HKK baseline since the peak coincides with the second oscillation 
maxima and the L/E is different the degeneracy pattern is somewhat different. 
In this case, as can be seen from the first plot in the second row, for 
neutrinos the NH probability can have hierarchy degeneracy only 
in the range  $ -30^\circ $ to $ 90^\circ $ excepting the range
$ 10^\circ < \dcp < 60^\circ $ for LO (green band) 
for which there is no degeneracy.   
For IH, on the other hand there is degeneracy over the entire 
range of $\dcp$. Similarly the plot in the right panel of the 2nd row 
exhibits that for NH there is hierarchy degeneracy over the whole range. 
While for IH the hierarchy degeneracy occurs 
in the range  $ -180^\circ $ to $ -120^\circ $ and
$ 130^\circ $ to $ 180^\circ$ barring the small range $-170^\circ < \dcp < -150^\circ$ for LO. 
From these discussions it is clear that for true NH, neutrino run is 
better for hierarchy sensitivity whereas for true IH, antineutrinos 
are more useful. 

For octant degeneracy if we check the NH plot in the left panel then 
(the green and yellow 
bands) we see that over most of the $\dcp$ range a horizontal line drawn through
these bands intersect the probabilities at $\dcp$ values belonging to 
different half planes giving rise to right-hierarchy -- wrong octant -- wrong 
$\dcp$ solutions. Over the range of $\dcp$ for which hierarchy degeneracy 
occurs there can also be wrong-hierarchy -- wrong octant  solutions 
in the same half plane of $\dcp$. For IH also, over almost full CP 
range octant degeneracy is seen from the probability curves. 
Similar conclusions can be drawn from the antineutrino probabilities. 
Note that for the Korean baseline and energy for neutrinos
the IH-LO band (red) lies above the IH-HO (blue) band unlike the 
other two experiments.  
For IH 
$\beta_3$ being greater than
$\beta_1$ the slope is negative and hence LO has a higher probability. 
Similarly for the antineutrinos NH-LO (green band) 
has a higher probability than NH-HO (yellow band). 
% for IH 
%$\beta_3$ being greater than
%$\beta_1$ the slope is negative and hence the behaviour of the probabilities
%with respect to octant is different
We can also see that the  
the curves for
LO and HO are much 
closer for the 1100 baseline. 
In addition, the widths of the octant bands are very small
which is reflective of the fact that the variation  
in the probability with octant is much less as we have seen earlier. 
Thus the octant sensitivity is 
expected to be less for this baseline and energy combination. 
From the equation \ref{eq:pmuenew},
we can write the difference of the probabilities 
for two different $\theta_{23}$ values as,  
\begin{equation}
P_{\mu e}(\theta_{23})-P_{\mu e}(\theta_{23}^{\prime}) \approx (\beta_{1}-\beta_{3})(\sin^2\theta_{23}-\sin^2\theta_{23}^{\prime})
\label{eq:th23diff} 
\end{equation}
Note that the $\beta_2$ term is not included since it $\sin2\theta_{23} \sim 1$ 
over the range in which $\theta_{23}$ is varied. 
Because of this the difference is independent of $\dcp$. This implies
the width of the $\theta_{23}$ band is expected to be the same 
for all values of $\dcp$ which can be seen from the figure \ref{prob-plots}. 
Inserting the values of $\beta_1$ and $\beta_3$ from table \ref{tab:beta} 
and putting $\theta_{23}$ as $39^\circ$ and $\theta_{23}$ 
as $42^\circ$ we find that the width of the  LO band as 
0.006 (DUNE), 0.005 (T2HK) and 0.002 (T2HKK). 
This is agreement with what is observed in the plots.

%Although DUNE has no hierarchy degeneracy still DUNE has significant octant degeneracy we can understand it from the bands, in fig(e) the yellow band is NH-HO and green band is NH-LO, these two bands are degenerate. In the same figure if we study the IH bands then the degeneracy worsens. Similar explanations can also be given for antineutrino probabilities.  

%%%%%%%%%%%%%%%%%%%%%%%%%%%%%%%%%%%%%%%%%%%%%%%%%%%%%%%%%%%%%%%%%%%%%%%%%%%%%%%%%%%%%%%%%%%%%%%%%%%%%%

\section{Experimental and Simulation details} 

In this section we elucidate briefly on the experimental set ups 
that have been used in our analysis. 

\subsubsection*{T2HK and T2HKK}
T2HK (Tokai-to-Hyper-Kamiokande) is a natural extension to the existing 
T2K experiment. It's default plan was to have two Hyper-Kamiokande detectors 
(cylindrical water tanks) of 187 kt at 295 km baseline in Kamioka.
T2HKK, an alternative to T2HK, is a newly proposed option 
which plans to have one of it's detectors at 295 Km in Kamioka mine 
and another at 1100 km in Korea. The water-cherenkov detector 
in Korea could be placed at one of the three suggested off-axis (OA) 
angles $1.5^\circ$, $2^\circ$ or $2.5^\circ$. The optimization of 
these OA angles to give maximum sensitivity to various neutrino 
oscillation parameters has been explored in ref. \cite{Abe:2016ero}.
This study indicates that the optimal configuration is to place 
the detector at $1.5^\circ$ OA angle. Therefore, we 
have considered the 187 kt Korean detector at an OA angle of $1.5^\circ$
in our simulations.
The proposed runtime for both configurations is 
1:3 in neutrino and antineutrino modes and the 
total exposure of $27\times 10^{21}$ Protons on Target(POT) 
which is obtained 
by a beam energy of 1.3 MW and 10 years of experiment run. 
%To simulate the experiments T2HK and T2HKK we have used the package 
%General Long Baseline Experiment Simulator (GLoBES) \cite{Globes:2004ka}. 
The detector specifications and the systematic uncertainties 
are consistent with the proposal given in ref. \cite{Abe:2016ero}. 
The simulations in this work are carried out after matching 
with the signal, background event spectra and the sensitivities 
presented in ref. \cite{Abe:2016ero}.
%and the mass hierarchy significance with 
%The background errors considered in our simulations are Neutral Current 
%background events, mis-identified $\nu_\mu$ events, intrinsic beam backgrounds 
%and wrong-sign {\bf{wrong sign what ?}} 
%backgrounds.
% The detector systematic uncertainties are parametrized in terms of 
%four nuisance parameters as follows : signal normalization error 3.84\% (3.83\%), signal tilt error 10\% (10\%), background normalization error 3.84\% (3.84\%) and background tilt error 10\% (10\%)~ for the appearance (disappearance) channel. 
%In our simulation the oscillation parameters we have considered are given in the table \ref{param_values}.

\subsection*{DUNE}
Deep Underground Neutrino Experiment (DUNE) is a promising upcoming
long-baseline neutrino oscillation experiment supported by Long-Baseline 
Neutrino Facility (LBNF). LBNF and DUNE facilities together will constitute 
a high intensity neutrino beam of 0.5-8 GeV energy, a near detector at Fermilab site, a 40 kt liquid argon time-projection chamber (LArTPC) as far detector 
1300 km away at Sanford Underground Research Facility (SURF), South Dakota.
Simulation of the DUNE experiment is done by considering a beam power of 
1.2 MW which results in a total exposure of $10\times 10^{21}$ Protons on Target(POT) for a 10 years experimental run. The $\nu$ : $\bar{\nu}$ runtime ratio 
for DUNE is considered as 1:1. The experimental specifications for our 
simulation of DUNE are taken from the ref. \cite{Acciarri:2015uup}.
%The signal error, background error and tilt are 
%taken as $2.5\%$, $10\%$ and $2.5\%$ respectively.

\subsection*{T2K \& NO$\nu$A}
NO$\nu$A \cite{Itow:2001ee} and T2K \cite{Itow:2001ee} are currently operative long-baseline experiments. 
The NO$\nu$A experiment has a baseline of 812 km. For this experiment muon-neutrinos are directed towards a far detector of fiducial volume 14 kt located at Ash River Minnesota from the NuMI beam facility at Fermilab. In our simulation we have considered the projected exposure for \nova which consists 
of 3 year $\nu$ and  3 year $\bar{\nu}$ runs and total exposure of $7.3\times10^{20}$ POT. 
The T2K experiment uses the muon-neutrino beam at JPARC facility. 
The beam is projected towards a detector of fiducial volume 22.5 kt 
located at Kamioka with a baseline of 295 km. 
We have considered a runtime of 4$\nu$+4$\bar{\nu}$ with a total exposure as $8\times10^{21}$ POT.

%%%%%%%%%%%%%%%%%%%%%%%%%%%%%%%%%%%%%%%%%%%%%%%%%%%%%%%%%%%%%%%%%%%%%%%%%
\section{Hierarchy, octant and CP discovery potential of T2HK,T2HKK and DUNE}
%%%%%%%%%%%%%%%%%%%%%%%%%%%%%%%%%%%%%%%%%%%%%%%%%%%%%%%%%%%%%%%%%%%%%%%
%\section{Results}

In this section, we present a comprehensive analysis of  
the  hierarchy, octant and CP discovery sensitivity of the experiments
T2HK, T2HKK and DUNE with their proposed configurations. 
In addition we also present the sensitivities 
including the projected run times of T2K and \nova.
For performing the numerical simulation we have used the package 
General Long Baseline Experiment Simulator (GLoBES) \cite{globes1,globes2}.   
Our presentation facilitates a comparison of the performances 
of the different experiments.
We perform a $\chi^2$ test where, 
\begin{eqnarray}
 \chi^2_{{\rm stat}} = 2 \sum_i \lbrace N_i^{{\rm test}}-N_i^{{\rm true}}+N_i^{{\rm true}} ln \frac{N_i^{{\rm true}}}{N_i^{{\rm test}}} \rbrace, 
 \label{mh-eq}
\end{eqnarray} 
where $(N)_i^{{\rm true}}$ corresponds to the simulated data 
and  $(N)_i^{{\rm test}}$ is the number of events predicted by the 
theoretical model.  
The systematic errors were incorporated in terms of pull variables. 
We 
%have added a prior on $\theta_{13}$ and 
have marginalized 
over the test parameters. 
The true values of the parameters and the marginalization ranges of 
the other parameters are as given in Table 
\ref{param_values}.

%In a three flavour framework there can be two possible ordering
%of the mass eigenstates $m_i$. If $m_1 < m_2 < m_3$ then
%we get the normal hierarchy (NH) and if $m_3 > m_2 \approx m_1$ then
%it is called the inverted hierarchy (IH).
%Octant of $\theta_{23}$ refers to whether $\theta_{23} < 45^\circ$
%i.e it is in lower octant (LO)
%or if $\theta_{23} > 45^\circ$ i.e it lies in higher octant (HO).
%For $\dcp$ the values 0 and $\pm 180^\circ$ correspond to CP conservation.

We present the results for four true  
hierarchy-octant combinations --  NH-LO, NH-HO, IH-LO \& NH-HO.  
The true value of $\theta_{23}$ for LO is considered as $42^\circ$  
while for HO it is taken as $48^\circ$ throughout our analysis unless 
otherwise mentioned.  
All the figures in this section show 
the performances of T2HK, T2HKK and DUNE and their 
corresponding sensitivities are plotted in red, blue and green
colors respectively. 
The solid lines represent the individual sensitivities of these experiments
and the dashed lines correspond to the combined significance of the 
concerned experiment along with T2K and NO$\nu$A data.

%%%%%%%%%%%%%%%%%%%%%%%%%%%%%%%%%%%%%%%%%%
\subsection*{\bf Octant Sensitivity} \label{sec:octs-sens}
%%%%%%%%%%%%%%%%%%%%%%%%%%%%%%%%%%%%%%%%%
To resolve the octant ambiguity of $\theta_{23}$ is one of the important goals 
of current and up-coming LBL experiments. This gained more prominence
in the light of the tension between the recent results of T2K and NO$\nu$A 
on the octant of $\theta_{23}$ where the former favours near maximal 
value of $\theta_{23}$ and the latter excludes maximal $\theta_{23}$ with 
$2.6\sigma$ significance. 

\begin{figure}[h!]
%\vspace{-1.4cm} 
        \begin{tabular}{clr}
                \hspace*{-0.65in} 
                &
				\includegraphics[width=0.5\textwidth]{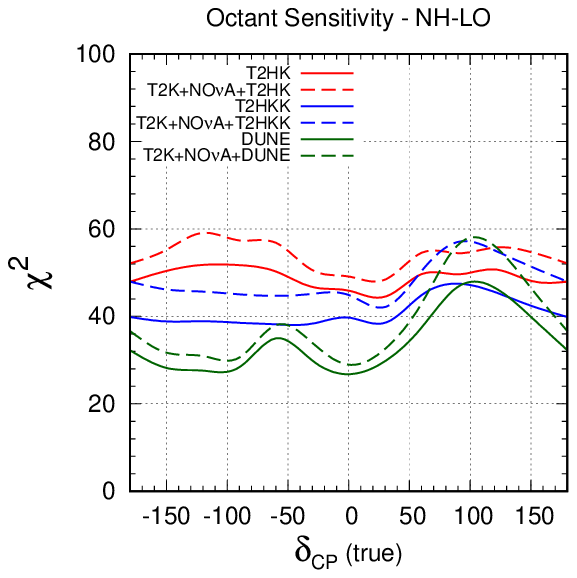}
				 \hspace*{-1.0in}
				\includegraphics[width=0.5\textwidth]{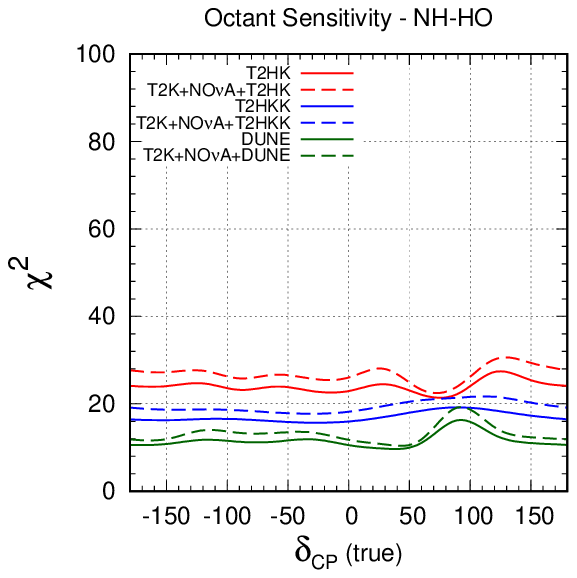} \\
				&
				\includegraphics[width=0.5\textwidth]{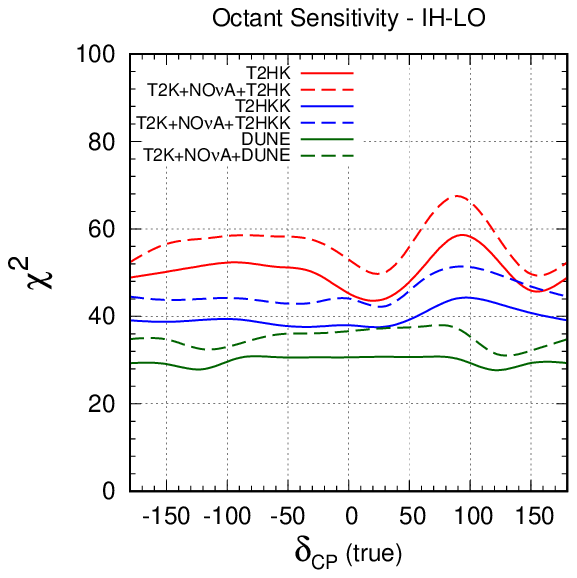}
				 \hspace*{-1.0in}
				\includegraphics[width=0.5\textwidth]{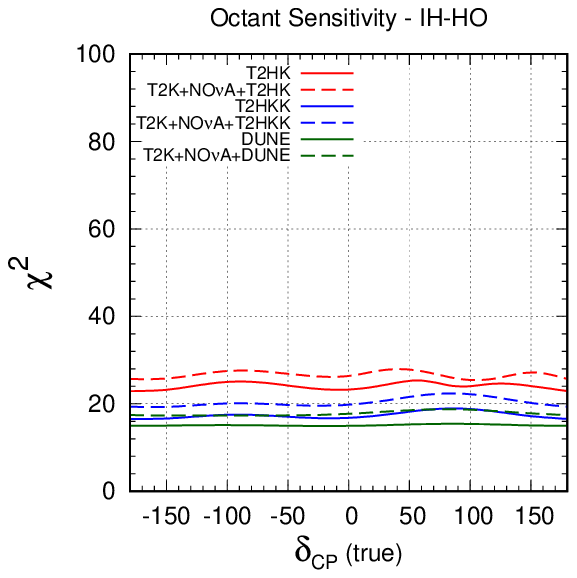}
		\end{tabular}
		\caption{Octant sensitivity in DUNE, T2HK and T2HKK with all hierarchy-octant configurations, first(second) row represent NH(IH) and first(second) columns represent LO(HO)}
		\label{octant-all-expt}
\end{figure}

However, the upcoming LBL experiments
T2HK, T2HKK and DUNE can overcome the various degeneracies that 
are effecting the octant sensitivity. 
The figure \ref{octant-all-expt}
shows the ability of the various configurations to exclude the wrong octant
of $\theta_{23}$ plotted as a function of true $\delta_{CP}$. 
Octant sensitivity is calculated by assuming a true octant in data and
considering the opposite octant as test in the theory. 
Marginalization is done over $\theta_{13}$, $\theta_{23}$(over the range of opposite octant), $|\Delta m^2_{31}|$, hierarchy and $\dcp$. 

We find that the sensitivity is higher for a true lower octant 
for all the experiments. 
There are two reasons for this.  Firstly for non-zero $\theta_{13}$ 
the $\chi^2$ vs $\theta_{23}$ curve is not symmetric about $45^\circ$. 
The two degenerate solutions in opposite octants are approximately 
related by $\theta_{23}^{LO} = 91.5^\circ - \theta_{23}^{HO}$ \cite{spuriousth23}. 
This implies that for a true $\theta_{23}$ of $42^\circ$ the degenerate 
minima will be at $49.5^\circ$ while for a true $\theta_{23}=48^\circ$ 
it will be at $43.5^\circ$. Thus the numerator in eq. \ref{mh-eq} is 
higher for a  true LO. 
On the other hand the denominator for the case of LO will be lower because,
in general, LO has lower probabilities compared to HO
and hence for LO the number of events will be reduced. 
Thus, the octant sensitivity is higher for LO 
%and octant  
%sensitivity will be reduced 
%\footnote{Note that for T2HKK LO probability 
%is higher for IH and neutrinos and NH and antineutrinos as is discussed
%in the previous section. However these are not the dominant probabilities 
%for these cases and hence this does not make any difference}.  
For a true LO 
greater than $5\sigma$ sensitivity is obtained for all the experiments. 
For true HO, only T2HK experiment can attain $5\sigma$ octant sensitivity. 
In general, the best octant sensitivity is achieved by the proposed T2HK
experiment for all the cases.  
The octant sensitivity for T2HKK experiment is less as compared to T2HK,
since as is evident from the probability discussions the 
octant sensitivity gets reduced for the 1100 km baseline.  
%This is expected from the probability behaviour as in fig. \ref{prob-plots}, because we can see that the 295 km baseline has good octant sensitivity whereas the 1100 km baseline has extremely large octant degeneracy. DUNE has also a good octant sensitivity, but DUNE has less sensitivity compared to T2HK because T2HKK has better detectors hence large statistics. 
%We can conclude that T2HK proposal is supposed to be the best experiment to resolve the octant degeneracy.
Addition of the data of T2K and NO$\nu$A  helps to 
increase the 
the significance by approximately 15\% in all cases.  
The octant sensitivity of DUNE is seen to be slightly less as compared to 
T2HK and T2HKK in most of the parameter range. This can be attributed to the 
lesser detector volume of DUNE. The underlying physics issues behind octant sensitivity of DUNE has been discussed in detail in \cite{Nath:2015kjg}. 
     
%\end{itemize}

%%%%%%%%%%%%%%%%%%%%%%%%%%%%%%%%%%%%%%%%%%%%%%%%%%%%%%%%%%%%%%%%%
\subsection*{\bf {Mass Hierarchy Sensitivity}} \label{mass-sub-sec}
\begin{figure}[h!]
%\vspace{-1.4cm} 
        \begin{tabular}{clr}
                \hspace*{-0.65in} 
                &
				\includegraphics[width=0.5\textwidth]{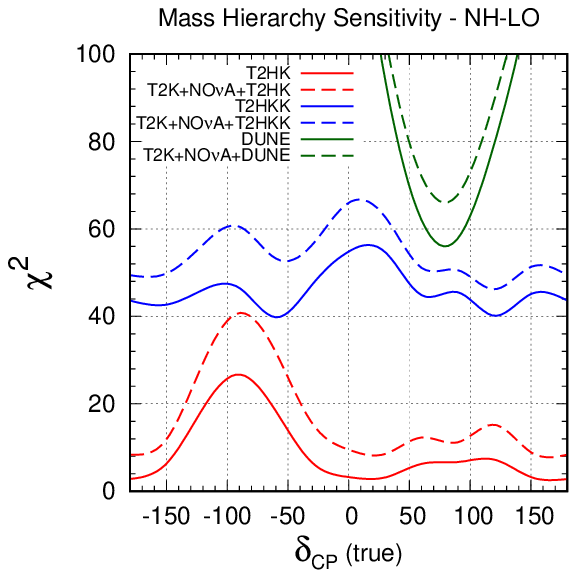}
				 \hspace*{-1.0in}
				\includegraphics[width=0.5\textwidth]{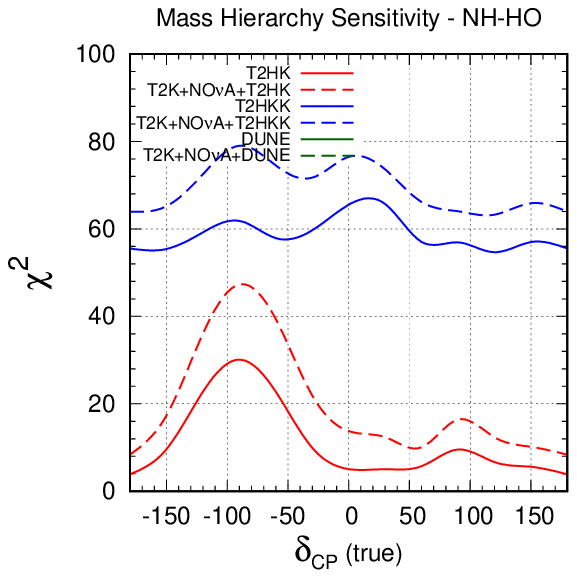} \\
				&
				\includegraphics[width=0.5\textwidth]{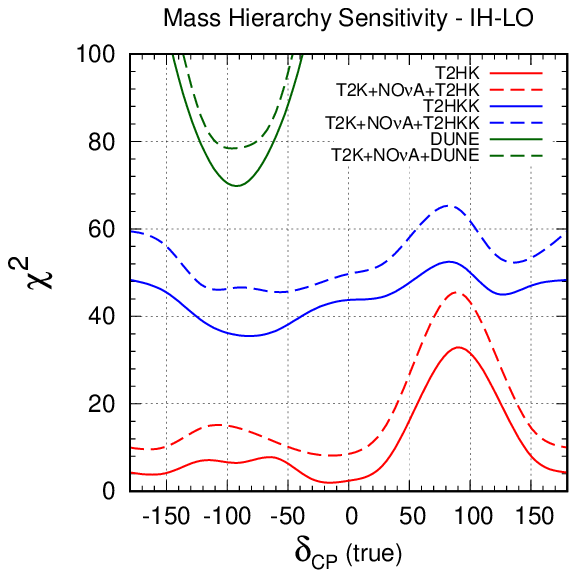}
				 \hspace*{-1.0in}
				\includegraphics[width=0.5\textwidth]{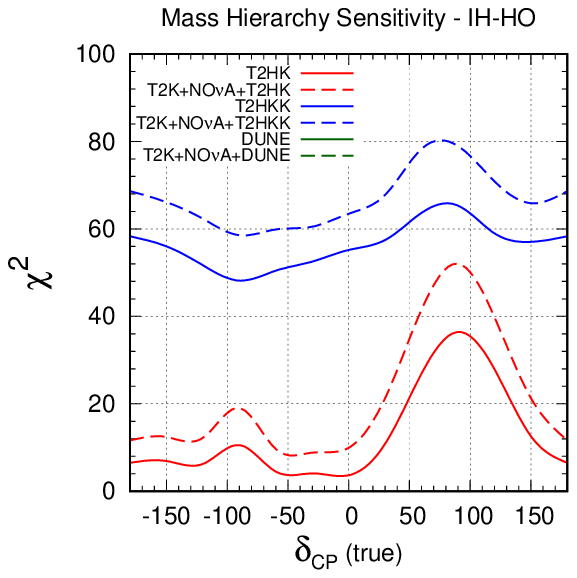}
		\end{tabular}
		\caption{Mass hierarchy sensitivity in DUNE, T2HK and T2HKK with all hierarchy-octant configurations, first(second) row represent NH(IH) and first(second) columns represent LO(HO)}
		\label{mass-all-expt}
\end{figure}

%%%%%%%%%%%%%%%%%%%%%%%%%%%%%%%%%%%%%%%%%%%%%%%%%%%%%%%%%%%%%%%%%%%%%%
Mass hierarchy sensitivity is calculated (using eq. (\ref{mh-eq})) by assuming 
a true hierarchy in the data and testing it by fixing the opposite hierarchy 
in theory. 
Marginalization is done over $\theta_{13}$, $\theta_{23}$, $|\Delta m^2_{31}|$ and $\dcp$. 
The figure \ref{mass-all-expt} demonstrates the mass hierarchy 
sensitivity of the proposed experiments. 
It is seen that for T2HK the 
hierarchy sensitivity is much reduced in the UHP for NH and in the LHP 
for the IH. This is due to the presence of hierarchy degeneracies
between NH-LO and IH-LO (green and red band) 
as can be seen from fig. \ref{prob-plots}.  
Similarly for IH, the unfavourable zone for hierarchy determination is the 
LHP.  Note that the WH-WO solutions are removed because of 
combined neutrino antineutrino run. 
% For instance 
%when we assume NH-LO (green band) as the true configuration one can see from 
%the top left plot of fig. (\ref{prob-plots}) 
%that there exists WH-RO and WH-WO solutions occurring with the 
%and red bands in the UHP. 
%In the case of NH-HO 
%one can see by drawing a horizontal line such that the yellow band (UHP)
%is degenerate with NH-LO (green),
% IH-HO (blue) in the LHP of $\dcp$. Similar conclusions can be drawn 
% for the other configurations from the probability plot of fig. (\ref{prob-plots}).
For T2HKK the hierarchy 
sensitivity is much higher.  
In this case the detectors are at two different baselines hence oscillation effects at both baselines will contribute. 
From the mass hierarchy plots fig. \ref{mass-hier1} we can conclude that there had been a significant increment in the overall mass hierarchy sensitivity over the previous case of both detectors being at 295 km baseline. 
In particular the degeneracy faced by the T2HK setup in the 
unfavourable region of $\dcp$ can be resolved by moving one 
detector to Korea.   
This is because for the Korean detector
the behaviour of the probability near 
second oscillation maxima shows 
that the degeneracy for neutrinos (NH) and for antineutrinos 
(IH) occur 
only over a small range of $\dcp$ values as discussed in the earlier section. 

%Since for NH the neutrino probability is higher, this dominates the 
%sensitivity
%and removes the degeneracy faced at the 295km baseline and hence 
%the sensitivity improves in the unfavourable zone compared to T2HK 
%experiment.       
%This is because for 1100 km baseline NH and neutrinos and IH and antineutrinos 
%have degeneracy only over a small range of $\dcp$ values. 
% because of the fact that the 1100 km baseline is not plagued with much hierarchy degeneracy 
For DUNE because of matter effect the hierarchy 
degeneracy is lifted. 
DUNE has very high hierarchy sensitivity and for HO the $\Delta \chi^2$ 
value is $>100$.  
Inclusion of T2K and \nova results can enhance the sensitivity by
 $\approx 10\%$ 
in all the three cases.

\subsection*{\bf CP Sensitivity}
\begin{figure}[h!]
%\vspace{-1.4cm} 
        \begin{tabular}{clr}
                \hspace*{-0.65in} 
                &
				\includegraphics[width=0.5\textwidth]{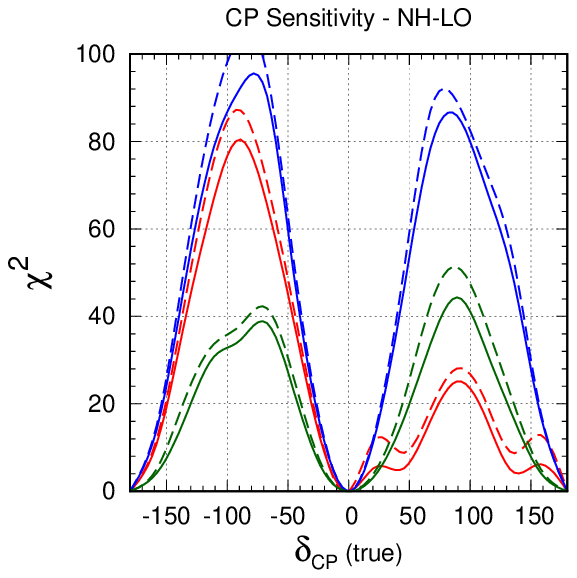}
				 \hspace*{-1.0in}
				\includegraphics[width=0.5\textwidth]{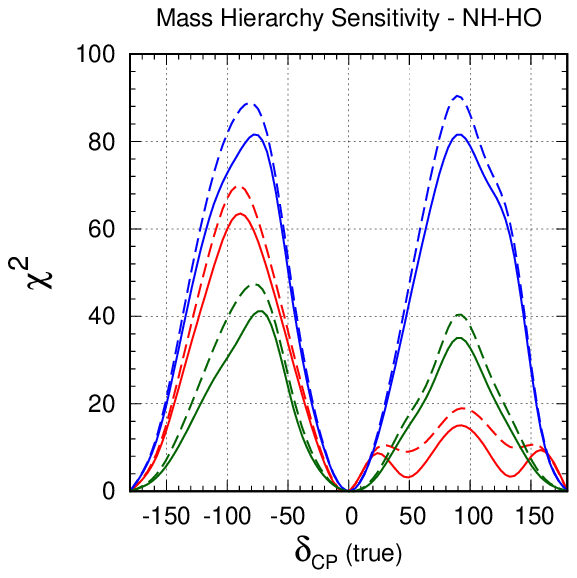} \\
				&
				\includegraphics[width=0.5\textwidth]{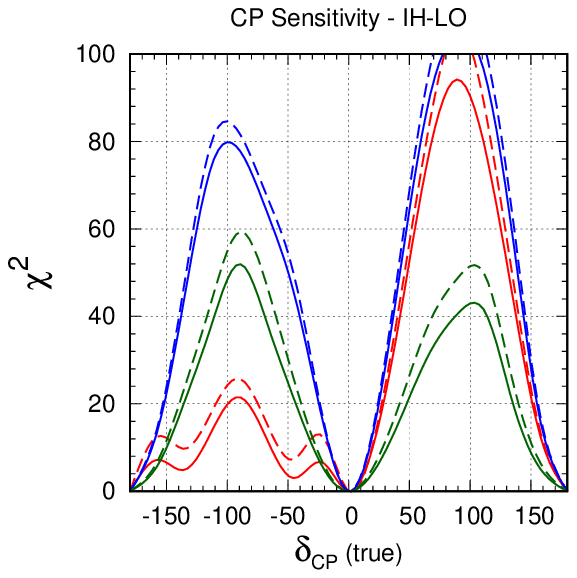}
				 \hspace*{-1.0in}
				\includegraphics[width=0.5\textwidth]{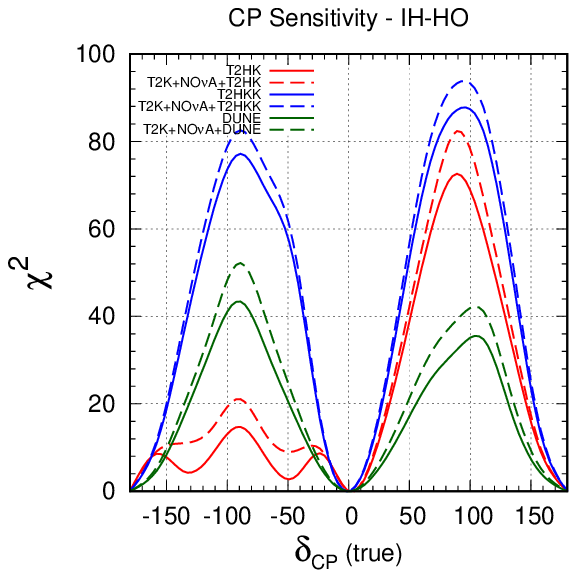}
		\end{tabular}
		\caption{CP sensitivity in DUNE, T2HK and T2HKK with all hierarchy-octant configurations, first(second) row represent NH(IH) and first(second) columns represent LO(HO)}
		\label{cp-all-expt}
\end{figure}

%Another goal of the oscillation experiments is to search for CP violation sensitivity. 
Calculation of CP discovery sensitivity is performed by simulating the 
data for all   
true $\dcp$ values and comparing them with CP conserving values 
$\dcp = -180^\circ, 0^\circ \&  180^\circ$. 
Marginalization is done over  $\theta_{13}$, $\theta_{23}$, $|\Delta m^2_{31}|$
and hierarchy. 
The CP discovery potential of the experiments under consideration 
is shown in figure \ref{cp-all-expt}. The maximum CP  discovery sensitivity 
can be achieved for $\dcp=\pm 90^\circ$.  For T2HK the CP discovery potential 
is seen to be less in one of the half-planes of $\dcp$. 
This is due to the presence of wrong hierarchy solutions. 
T2HKK can overcome this because it has a better 
hierarchy sensitivity and hence  the hierarchy-$\dcp$ degeneracy is resolved. 
In general T2HKK has a better CP sensitivity because at the second oscillation 
maxima the CP effect is larger and can compensate for the loss in flux 
due to a higher baseline \cite{Abe:2016ero}. In ref. \cite{Abe:2016ero} 
it  was shown that the difference in the  CP asymmetry between CP conserving 
and CP violating values is  more for the 1100 km baseline.
At the $\chi^2$ level this gets reflected in the tension between the 
neutrino and the antineutrino contribution to the $\chi^2$. 
For the T2HK experiment at 295 km the oscillation peak and the flux 
peak coincide and the probability for $\dcp=0$ and $\pm 180^\circ$ 
are equidistant from the probability at $\pm 90^\circ$. 
However for the 1100 km baseline as can be seen from the probability plot 
for either NH or IH as the true hierarchy, for neutrinos $\dcp = -90^\circ$ 
is closer to $\pm 180^\circ$ while for antineutrinos $\dcp = 0$ 
is closer to $\dcp = -90^\circ$. This creates a tension between the 
neutrino and the antineutrino $\chi^2$  which gives a better sensitivity to 
T2HKK as one of its baselines is at 1100 km. 
For DUNE, the wrong hierarchy solutions get resolved and hence the 
sensitivities do not suffer a drop in one half plane of $\dcp$ as in T2HK. 
In this case also, the 
neutrino and antineutrino tension can enhance the overall
$\chi^2$ for CP violation \cite{Nath:2015kjg}. 
However, since the statistics of DUNE is lower compared to T2HKK and T2HK 
it's CP discovery sensitivity is lower.

\section{Optimal Exposures}
{\label{optimal}} 

In the earlier section we have delineated the sensitivities of the 
different experiments and compared the performances using the 
proposed plans.  
These being very high statistics experiments in some 
cases very high sensitivity is seen to be achieved. 
In this section 
we present the optimum exposure that is needed by each experiment 
to attain $5\sigma$ significance for hierarchy and octant sensitivity. 
We also give the exposure required for achieving $3\sigma$ sensitivity 
for 60\% values of $\dcp$.  
We furnish our results for two representative hierarchy-octant 
configurations NH-LO and IH-HO. 
%These correspond approximately to the best and the worst cases of 
%hierarchy and octant sensitivity 
%-- as is seen from the previous section. 
For true LO and HO we have taken representative values of
$\theta_{23}$ as $42^\circ$ and $48^\circ$. 
%{{\bf Are these plots with 1:3 for T2HKK and 5+5 for dune ? }} 
 
\begin{figure}[h!]
                \hspace*{-0.35in} 
                \includegraphics[width=0.5\textwidth]{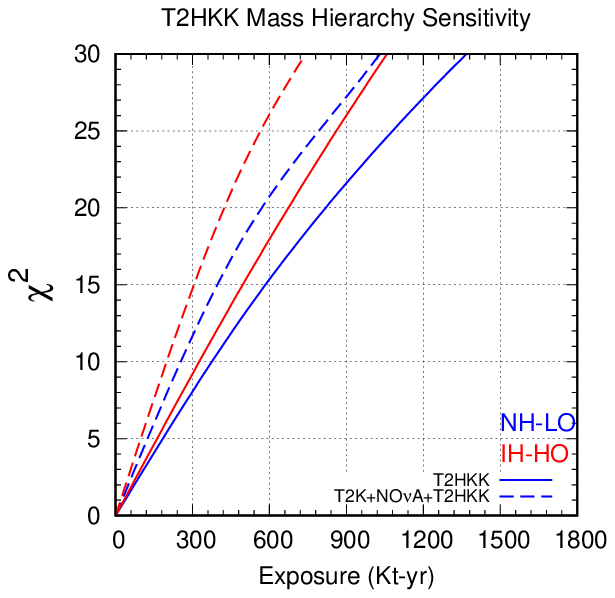}
               \hspace*{-1.0in}
                \includegraphics[width=0.5\textwidth]{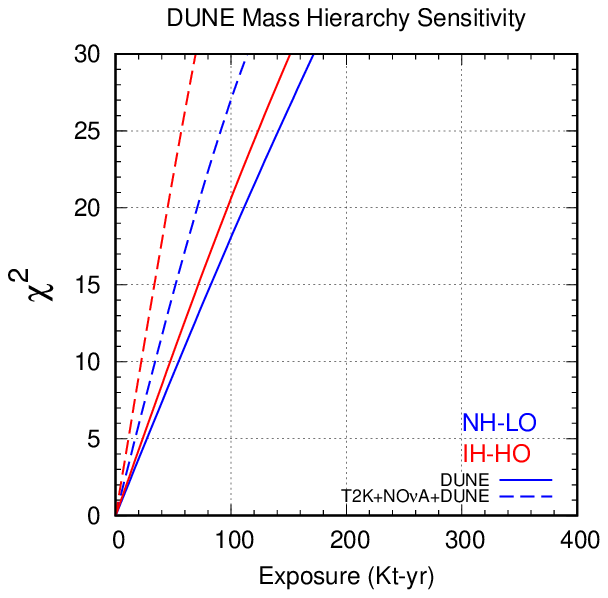}              
\caption{Mass hierarchy Sensitivity in T2HKK and DUNE versus exposure for true NH-LO with $\theta_{23} = 42^\circ$}
\label{runvary}
\end{figure}

In fig. \ref{runvary} we plot two sets of mass hierarchy
 $\chi^2$ vs 
exposure in kt-yr
for T2HKK and DUNE in the left and the right panels respectively. 
The blue (red) solid lines correspond to the sensitivity of 
true NH-LO (IH-HO) configuration of only the
concerned experiment.
The dashed lines represent the same when 
added with 3$\nu$ + 3$\bar{\nu}$ NO$\nu$A and 4$\nu$ + 4$\bar{\nu}$ T2K runs.
We marginalized over
 true $\delta_{CP}$ in the range $[-180^\circ,180^\circ]$ .
%The 
%red solid lines represent the sensitivity of only the concerned experiment 
%and the blue dashed lines represent the same when 
%added with 3$\nu$ + 3$\bar{\nu}$ NO$\nu$A and 4$\nu$ + 4$\bar{\nu}$ T2K runs.
%We have chosen the true combinations as IH-LO and marginalized over
%true $\delta_{CP}$ in the range $[-180^\circ,180^\circ]$ .
%As can be seen from the 
%fig. \ref{mass-all-expt} hierarchy sensitivity is poorer for LO and true NH and true IH
%gives similar sensitivity. 

From the left panel in fig. \ref{runvary} we observe that
considering only T2HKK the optimal exposure for $5\sigma$ hierarchy 
sensitivity is $\sim$ 1080 kt-yr for NH-LO and 680 kt-yr for IH-HO.
 This corresponds to volume of one detector. 
Therefore the total exposure for two detectors is 2160 kt-yr for NH-LO
and 1360 kt-yr for IH-HO. 
For the fiducial volume of 187 kt for a single detector in true NH-LO
configuration, this 
corresponds to  $\approx$ 6 years run time {\it{i.e.}} 
1.5 years in neutrino and 
4.5 years in antineutrino mode. The more optimistic case would be 
with true IH-HO where it requires only 3.6 years (0.9$\nu$+2.7$\bar{\nu}$)
of run time.
% for a single detector.
After adding T2K and \nova information this exposure for each detector
 is reduced to 
840 kt-yr for true NH-LO and 430 kt-yr for true IH-HO 
which for the proposed detector volume correspond to 4.5 years
and 2.3 years of run time respectively.  
Performing the similar analysis for DUNE we see that,
a minimum $5\sigma$ significance can be obtained for 140 kt-yr in true NH-LO. 
This corresponds to a run time of 3.5 years (split equally in neutrinos
and antineutrinos) for 40 kt detector volume. 
Adding the information from T2K and \nova reduces it to 70 kt-yr corresponding 
to just 1.75 years run time for DUNE. For the case where IH-HO   is 
the true combination $5\sigma$ significance can be obtained with an
exposure of 125 kt-yr, which reduces to 60 kt-yr once the data
from T2K and \nova is added. 
In  the first line of table \ref{opt-exp} we summarize the exposures for the different setups for $5\sigma$ hierarchy sensitivity  
for the favourable case of true IH-HO. 

%The most optimistic case among our representative
%configurations of true NH-LO and IH-HO to obtain a $5\sigma$ significance 
%with T2HKK and DUNE is IH-HO and the corresponding details of the exposures 
%are summarized in the table \ref{opt-exp}

%%%%%%%%%%%%%%%%%%%%%%%%%%%%%%%%%%%%%%%%%%%%%%%%%%%%%%%%%%%%%%%%%%%%%%

Similarly we also find out the optimum exposure for 
$5\sigma$ octant sensitivity. For this case we present our results 
for T2HK and DUNE.  Since the
octant sensitivity of T2HKK is poorer as compared to 
T2HK, as seen in the  earlier section \ref{sec:octs-sens}, we  
present the minimal optimal exposure required for T2HK to 
resolve the octant of $\theta_{23}$ with 
$5\sigma$ sensitivity assuming true NH-LO and IH-HO.
%In this case the representative values for true LO and HO  
%are taken to be $\theta_{23} = 42^\circ$ and $\theta_{23} = 48^\circ$ 
%and true $\delta_{CP}$ is marginalized in the range $ [-180,180]$.

\begin{figure}[h!]
                \hspace*{-0.35in} 
                \includegraphics[width=0.5\textwidth]{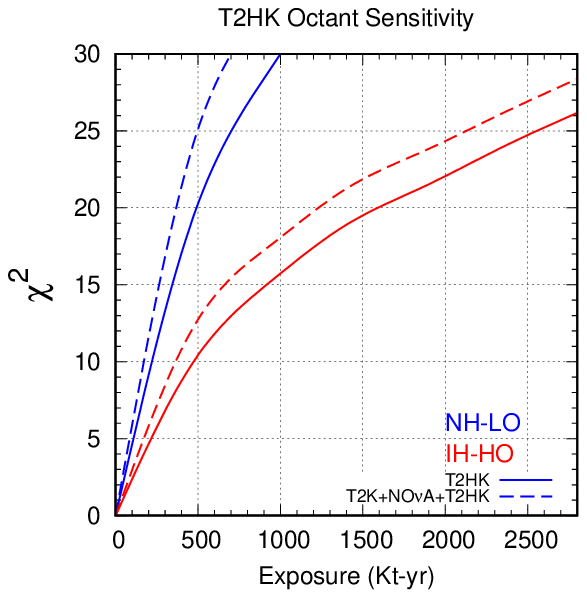}
                \hspace*{-1.0in}
                \includegraphics[width=0.5\textwidth]{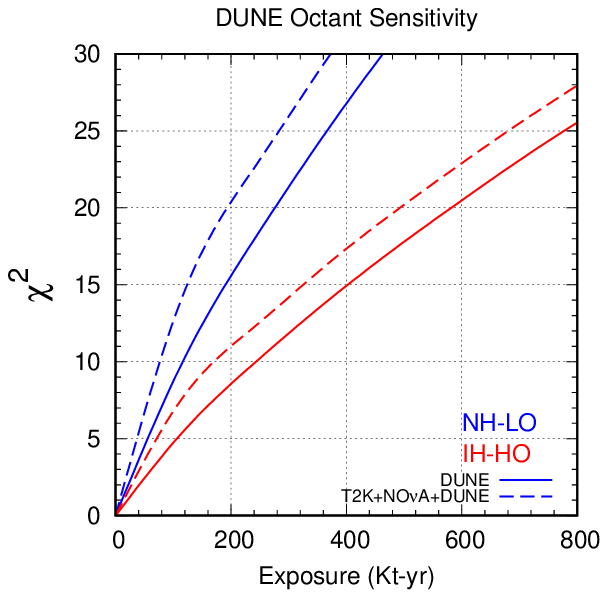}              
\caption{Octant Sensitivity in T2HK and DUNE versus exposure for true NH-LO}
\label{oct-runvary}
\end{figure}

We observe that overall $5\sigma$ significance can be obtained 
by the T2HK experiment for an exposure of 2500 kt-yr for one detector 
which corresponds to approximately 13.4 years of run time in IH-HO as
can be seen from the red solid line in fig. \ref{oct-runvary}.
Including the information from T2K and \nova the exposure reduces to 
2100 kt-yr corresponding to a run time of 11 years approximately. 
%{\bf{Note that for T2HKK this exposure is xxxx years.}} 
The second panel represents the octant sensitivity for DUNE. 
In this case, for true IH-HO (red curves) it can be seen that
 the exposure for $5\sigma$ octant sensitivity is 
800 kt-yr which with the proposed volume of 40 kt corresponds 
to a runtime of 20 years. Adding T2K and \nova information 
the optimal exposure is 680 kt-yr which corresponds to 17 years runtime. 
This is not surprising since as we have seen in the previous section
that DUNE doesn't reach $5\sigma$ octant sensitivity in 10 years for 
IH-HO with their proposed volume. 
However, from the blue lines of fig. \ref{oct-runvary}, we can see that 
the true hierarchy--octant configuration NH-LO gives 5$\sigma$ sensitivity 
for comparatively less exposure for both the experiments. We list the
exposures required for this optimistic case in table \ref{opt-exp}.
The number of operative years needed for T2HK to attain 5$\sigma$ 
octant sensitivity is 3.8 yr in its individual capacity and 2.7 yrs 
when the data from T2K and \nova experiments is added. However, DUNE 
requires 10 years of data taking to resolve octant of $\theta_{23}$ which 
reduces to 7.5 years when the T2K and \nova results are taken into account.

%%%%%%%%%%%%%%%%%%%%%%%%%%%%%%%%%%%%%%%%%%%%%%%%%%%%%%%%%%%%%%%%%%%%%
%%%%%%%%%%%%%%%%%%%%%%%%%%%%%%%%%%%%%%%%%%%%%%%%%%%%%%%%%%%%%%%%%%%%%%%%%%%

\begin{figure}[h!]
                \hspace*{-0.35in} 
                \includegraphics[width=0.5\textwidth]{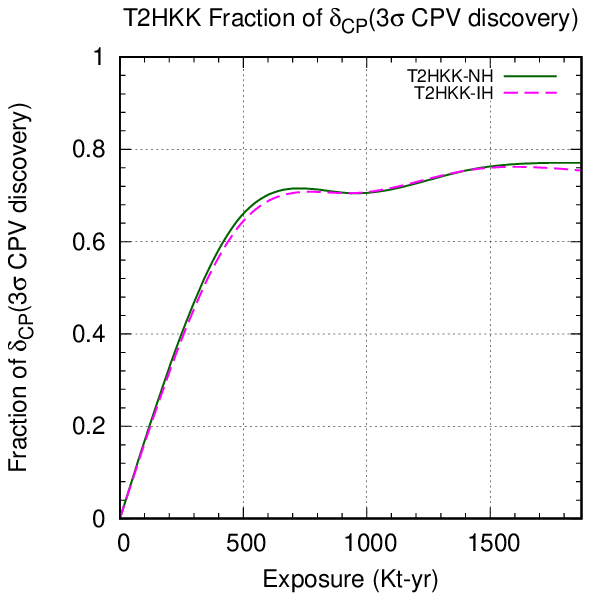}
                \hspace*{-1.0in}
                \includegraphics[width=0.5\textwidth]{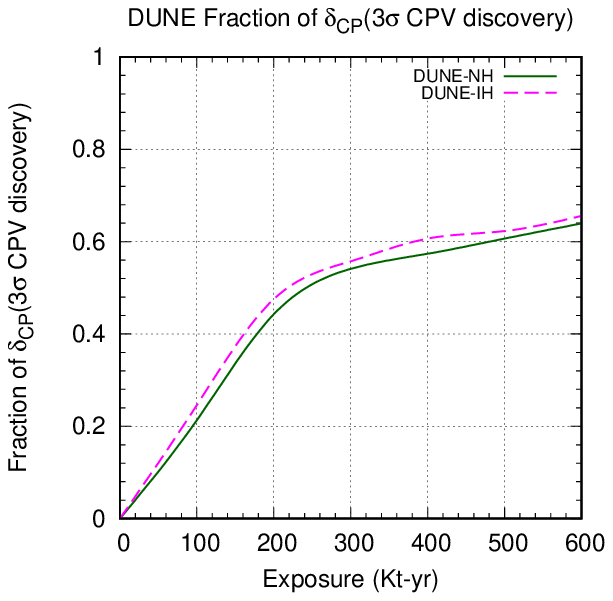}              
\caption{Fraction of $\dcp$ in T2HKK and DUNE versus exposure for $3 \sigma$ CPV sensitivity }
\label{cpd-runvary}
\end{figure}

%The other genre which the neutrino oscillations experiments are trying to look into is  CP violation sensitivity. 
In figure \ref{cpd-runvary} we display the fraction of $\dcp$(true) 
values for which
the experiments can observe CP violation with minimum $3 \sigma$ significance. 
We perform this study for both NH and IH.

For calculating the fraction of $\dcp$ values we compare the true $\dcp$ 
values against the CP conserving test $\dcp$ values of $0^\circ$ and $ \pm 180^\circ$.
The fraction of $\dcp$ for more than $3 \sigma$ CP 
discovery significance is the ratio of true $\dcp$ 
values for which the CP significance is more than $3 \sigma$ to the 
total number of $\dcp$ values.
 For true $\theta_{23}$ we 
consider three values $42^\circ$, $45^\circ$ and $48^\circ$ and 
choose the minimum $\chi^2$ amongst this set. 
We have also marginalized over $\theta_{13}$, $\theta_{23}$, $\Delta m^2_{31} $ 
and hierarchy in test. 
The fig. \ref{cpd-runvary} contains two plots for each 
experiment T2HKK(left) and DUNE(right). 
Each plot has two lines, green(solid) line represents
true-NH while magenta(dashed) represent true-IH. 
We see that 
60\% coverage in $\dcp$(true) values is obtained by an exposure of 400 kt-yr 
of T2HKK (each detector) both for NH and IH. However for DUNE
the exposure is 500 kt-yr for NH and 400 kt-yr for IH.
This corresponds to approximately 2 years of running of 
T2HKK and 12.5 years of running of DUNE with their proposed volumes.  

% As we know that the total proposed exposure for T2HKK is 1870 Kton-yr for each detector hence T2HKK can detect CP violation at lot less total exposure than the projected exposure. Now let us concentrate on the DUNE experiment, this experiment can provide a 40\% coverage with an exposure of 160 Kton-yr, but if we wish for 60\% coverage total exposure of 400 Kton-yr is required, with true hierarchy as NH this goal can be achieved but if true hierarchy is IH total coverage of 60\% cannot be achieved with an exposure below 500 Kton-yr. But, the proposed exposure for is 400 Kton-yr, so if true hierarchy is IH then DUNE will need to increase their exposure for better CP discovery potential. 

%%%%%%%%%%%%%%%%%%%%%%%%%%%%%%%%%%%%%%%%%%%%%%%%%%%%%%%%%%%%
\begin{table}
\begin{center}
\begin{tabular}{|c|c|c|c|c|c|c|}
\hline
Sensitivity   & T2HKK & T2HK & T2HKK+T2K+NO$\nu$A & T2HK+T2K+NO$\nu$A & DUNE & DUNE+T2K+NO$\nu$A  \\       
\hline
Hierarchy($\chi^2=25$)(IH-HO)         & 680  & --  &  430 & -- & 140  & 70  \\
%\hline
Octant($\chi^2=25$) (NH-LO)           & -- & 700   & --  &  500  & 400  & 300  \\
%\hline
CP(60$\%$ at $\chi^2=9$, NH)        & 400 & -- & --    &  --  & 500 & --   \\
CP(60$\%$ at $\chi^2=9$, IH)        & 400 & -- & --    &  --  & 400 & --   \\

\hline
\end{tabular}
\end{center}
\caption{Minimum exposures required for hierarchy, octant and CP in units of kt-yr}
\label{opt-exp}
\end{table}
%%%%%%%%%%%%%%%%%%%%%%%%%%%%%%%%%%%%%%%%%%%%%%%%%%%%%%%%%%%%%%%%%%%%%

%%%%%%%%%%%%%%%%%%%%%%%%%%%%%%%%%%%%%%%%%%%%%%%%%%%%%%%%%%%%
%\begin{table}
%\begin{center}
%\begin{tabular}{|c|c|c|c|c|c|c|}
%\hline
%Sensitivity   & T2HKK & T2HK & T2HKK+T2K+NO$\nu$A & T2HK+T2K+NO$\nu$A & DUNE & DUNE+T2K+NO$\nu$A  \\        
%\hline
%Hierarchy($\chi^2=25$)         & 1250  & --  &  800 & -- & 120  & 80  \\
%%\hline
%Octant($\chi^2=25$)            & -- & 2500   & --  &  2100  & 800  & 680  \\
%%\hline
%CP(60$\%$ at $\chi^2=9$, NH)        & 400 & -- & --    &  --  & 500 & --   \\
%CP(60$\%$ at $\chi^2=9$, IH)        & 400 & -- & --    &  --  & 400 & --   \\
%
%\hline
%\end{tabular}
%\end{center}
%\caption{Minimum exposures required for hierarchy, octant and CP in units of kt-yr}
%\end{table}
%%%%%%%%%%%%%%%%%%%%%%%%%%%%%%%%%%%%%%%%%%%%%%%%%%%%%%%%%%%%%%%%%%%%%

%%%%%%%%%%%%%%%%%%%%%%%%%%%%%%%%%%%%%%%%%%%%%%%%%%%%%%%%%%%%%%%%%%%%%%%%
\section{Neutrino and antineutrino run optimization in  T2HK \& T2HKK}
%%%%%%%%%%%%%%%%%%%%%%%%%%%%%%%%%%%%%%%%%%%%%%%%%%%%%%%%%%%%%%%%%%%%%%%%%%%%
The proposed total runtime for the  T2HK and T2HKK experiment is 10 years,
which will consists of 2.5 years  $\nu$ run and 7.5 years $\bar{\nu}$ run. 
This 1$\nu$ : 3$\bar{\nu}$ runtime ratio was chosen to keep the 
number of neutrino events comparable to that of the antineutrino events.
%, this happens because the cross sections for neutrino detection is thrice of that of the antineutrinos. 
%Do we really need the number of events to be comparable for better sensitivities?
In this section, we explore the possibilities of acquiring better sensitivities
by considering 3$\nu$ : 1$\bar{\nu}$ and 1$\nu$ : 1$\bar{\nu}$ 
as alternative runtime ratios.
%For a better understanding of the roles played by $\nu$ and $\bar{\nu}$ beams 
%we start by taking a runtime of 0$\nu$ + 10$\bar{\nu}$, after which we 
%increase $\nu$runs step by step while fixing the total runtime to 10 years.

\begin{figure}[h!]
%\vspace{-1.4cm} 
        \begin{tabular}{lr}
%                \hspace*{-0.35in} 
%                \includegraphics[width=0.5\textwidth]{chisq-true_nhlo-295km-all.eps}
%                & 
%                \hspace*{-1.0in}
%                \includegraphics[width=0.5\textwidth]{chisq-true_nhho-295km-all.eps}\\
%                \hspace*{-0.35in} 
%                \includegraphics[width=0.5\textwidth]{chisq-true_ihlo-295km-all.eps}
%                & 
%                \hspace*{-1.0in}
%                \includegraphics[width=0.5\textwidth]{chisq-true_ihho-295km-all.eps}\\
                &
                \hspace*{-0.65in} 
                \includegraphics[width=0.5\textwidth]{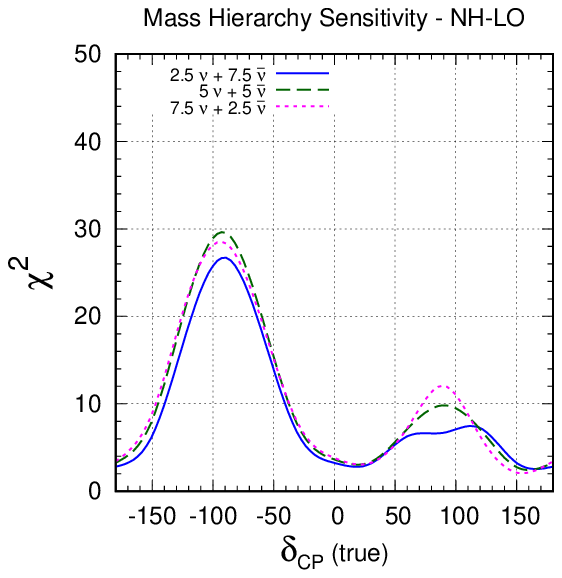} 
                \hspace*{-1.0in}
                \includegraphics[width=0.5\textwidth]{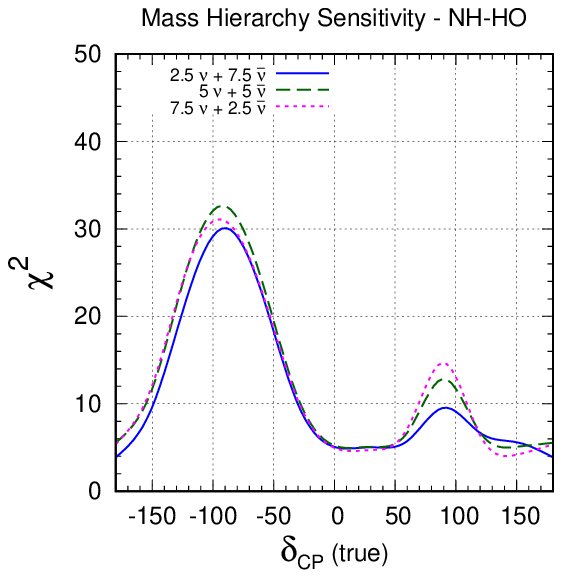}\\
                 &
                \hspace*{-1.0in}
                \includegraphics[width=0.5\textwidth]{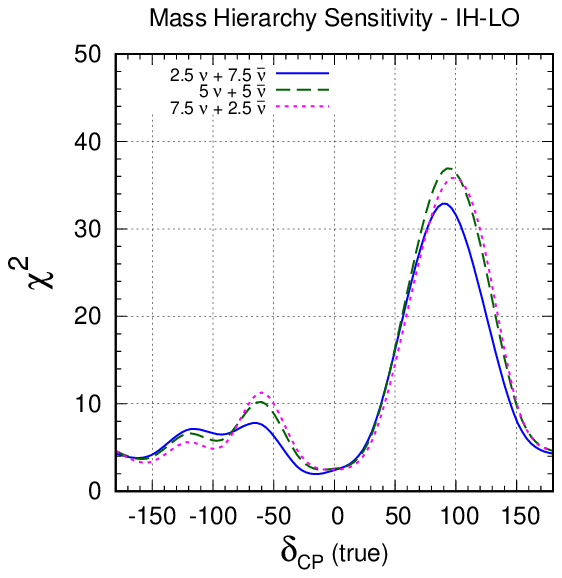}
               
                \hspace*{-1.0in}
                \includegraphics[width=0.5\textwidth]{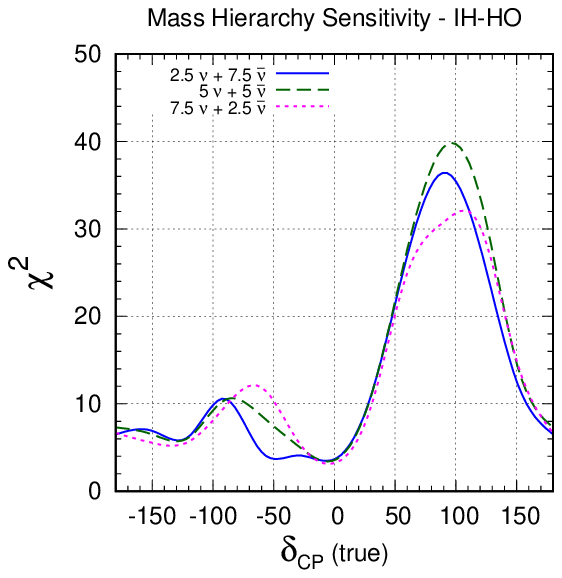}
\end{tabular}
\caption{Mass hierarchy $\chi^2$ vs $\dcp$ plots for T2HK (true NH first row and true IH second row). The labels signify the $\nu$+$\bar{\nu}$ runs.} 
\label{mass-hier}              
               \end{figure} 
\begin{figure}[h!]
%\vspace{-1.4cm} 
        \begin{tabular}{lr}
				 &	                 
                 \hspace*{-0.65in}
                \includegraphics[width=0.5\textwidth]{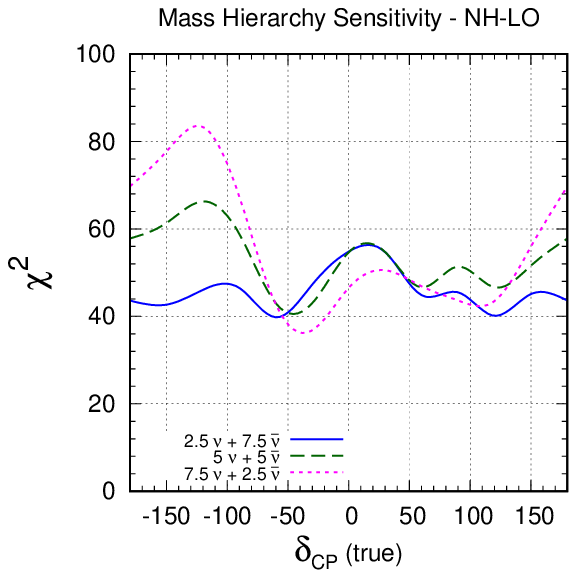}
                
                \hspace*{-1.0in}
                \includegraphics[width=0.5\textwidth]{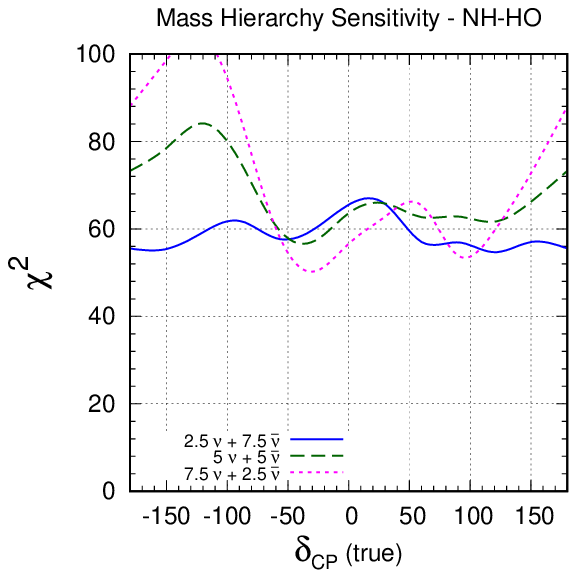}\\
                &
                \hspace*{-1.0in}
                \includegraphics[width=0.5\textwidth]{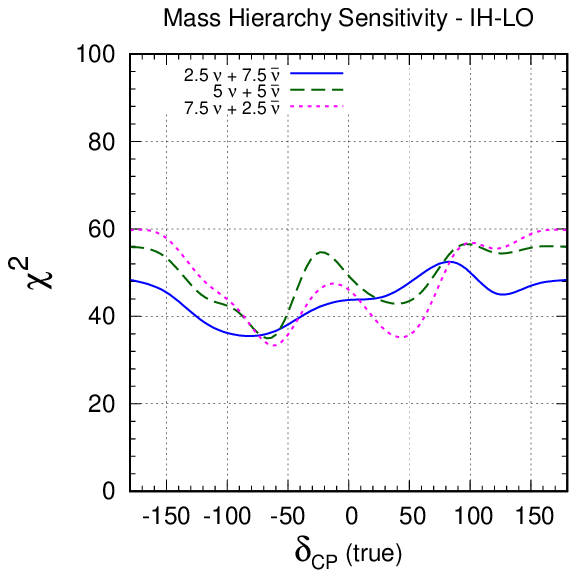}
                
                \hspace*{-1.0in}
                \includegraphics[width=0.5\textwidth]{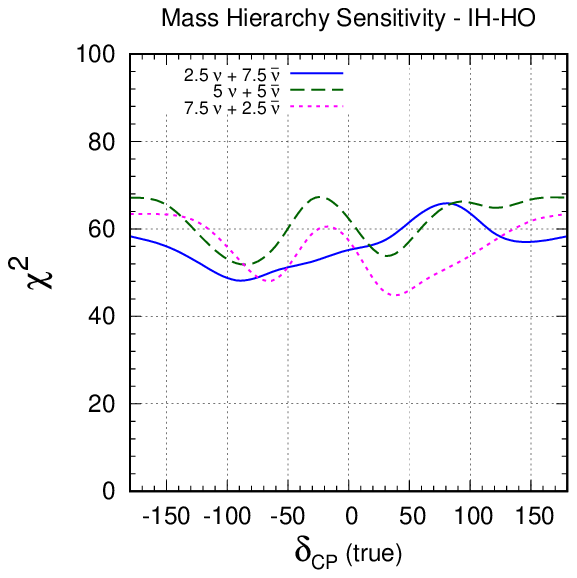}
        \end{tabular}
\caption{Mass hierarchy $\chi^2$ vs $\dcp$ plots for T2HKK (true NH first row and true IH second row). The labels signify the $\nu$+$\bar{\nu}$ runs.}   
\label{mass-hier1}
\end{figure}

\subsection{Analysis of Mass hierarchy sensitivity}
 
Fig. (\ref{mass-hier}) and fig. (\ref{mass-hier1}) represent the mass hierarchy 
sensitivities of T2HK, T2HKK.
In each plot three different run time ratios of 2.5$\nu$ + 7.5$\bar{\nu}$, 
7.5$\nu$ + 2.5$\bar{\nu}$, 5$\nu$ + 5$\bar{\nu}$ are shown by 
blue, magenta and green curves respectively. 
%The fig. (\ref{mass-hier}) represents $\chi^2$ vs $\dcp$ for T2HK \ref{mass-hier1}
%represent T2HKK i.e. one detector at 295 km and the other at 1100 km. 
From all the plots of fig. ((\ref{mass-hier}) corresponding to 
the true configurations -- NH-LO, NH-HO,
IH-LO and IH-HO we find that for T2HK all the three runtime ratios 
give hierarchy sensitivity in the same ballpark. 
The combinations $7.5+2.5$ and $5+5$ fare slightly better because 
of enhanced statistics except for the UHP of true IH-HO where $7.5+2.5$ 
is slightly lower than the other two cases. 
This is because in the UHP for IH-HO the neutrino probability
is impaired 
by WH-WO and RH-WO degeneracies unlike that in antineutrinos. Thus, 
the wrong octant solutions 
can be removed by considering more antineutrino runs as can be seen from
the UHP of true IH-HO configuration in the fig. \ref{mass-hier} where
the proposed 2.5$\nu$ + 7.5$\bar{\nu}$
gives better sensitivity than 7.5$\nu$ + 2.5$\bar{\nu}$.
However for T2HKK from fig. (\ref{mass-hier1}) we see that 7.5$\nu$ + 2.5$\bar{\nu}$ and 5$\nu$ + 5$\bar{\nu}$ give 
better sensitivity when compared to the proposed ratio of
2.5$\nu$ + 7.5$\bar{\nu}$ 
for some values of $\dcp$.  
If we compare the best two cases we infer that for 7.5$\nu$ + 2.5$\bar{\nu}$ 
the significance is quite high with respect to the significance 
of 2.5$\nu$ + 7.5$\bar{\nu}$ both in upper and lower half plane, 
except the region with $-20^\circ < \dcp < 40^\circ$ where 2.5$\nu$ + 7.5$\bar{\nu}$ 
gives better hierarchy sensitivity. The greater hierarchy sensitivity in T2HKK 
can be understood from fig. \ref{prob-plots} which depicts the oscillation probabilities. 
For 1100 km baseline in the region with $\dcp < -20^\circ$ and $\dcp > 40^\circ$ 
the neutrino appearance probabilities are not degenerate w.r.t. hierarchy, 
hence sensitivity in this region is governed by neutrino appearance, 
in the region $-20^\circ < \dcp < 40^\circ$  the sensitivity is governed 
by antineutrino appearance because in this region neutrino appearance probability
is degenerate but antineutrino appearance is non-degenerate. 
Considering both baselines the IH sensitivity behaves differently 
because of the non-degenerate behaviour of the probabilities at 1100 km baseline.
 Similar to NH we obtain better sensitivities at the regions $\dcp < -20^\circ$ and $\dcp > 40^\circ$ for 7.5$\nu$ + 2.5$\bar{\nu}$ and $-20^\circ < \dcp < 40^\circ$ for 2.5$\nu$ + 7.5$\bar{\nu}$.
We can conclude from this discussions that in the regions of 
maximum CP violation  5$\nu$+5$\bar{\nu}$ give somewhat better sensitivity for 
T2HKK experiment. 

 %$\theta_{23}$(test) for NH-LO at $\delta_{CP}(true) =-90^\circ$ 295km + 1100km}

\subsection{Analysis of Octant sensitivity}

\begin{figure}[h!]
%\vspace{-1.4cm} 
        \begin{tabular}{lr}
				 &	                 
                 \hspace*{-0.65in}
                \includegraphics[width=0.5\textwidth]{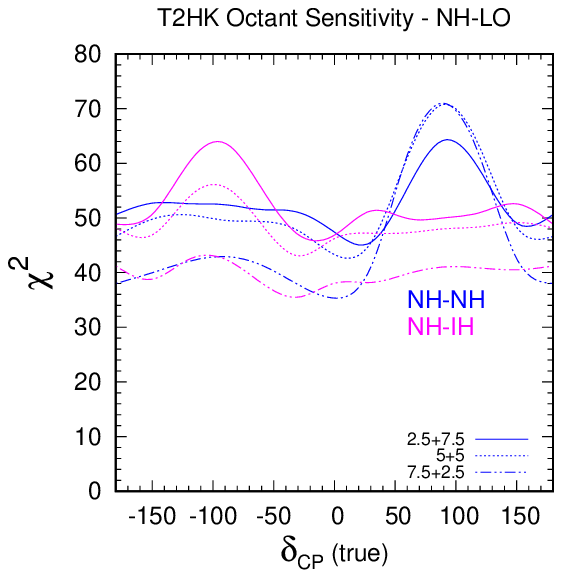}
                
                \hspace*{-1.0in}
                \includegraphics[width=0.5\textwidth]{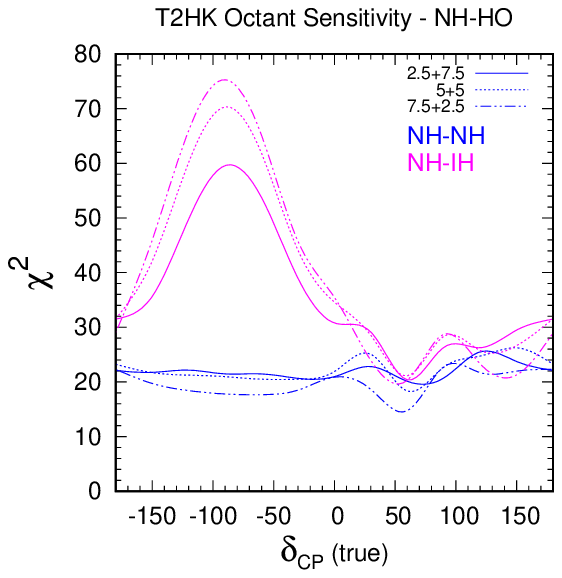}\\
                &
                \hspace*{-1.0in}
                \includegraphics[width=0.5\textwidth]{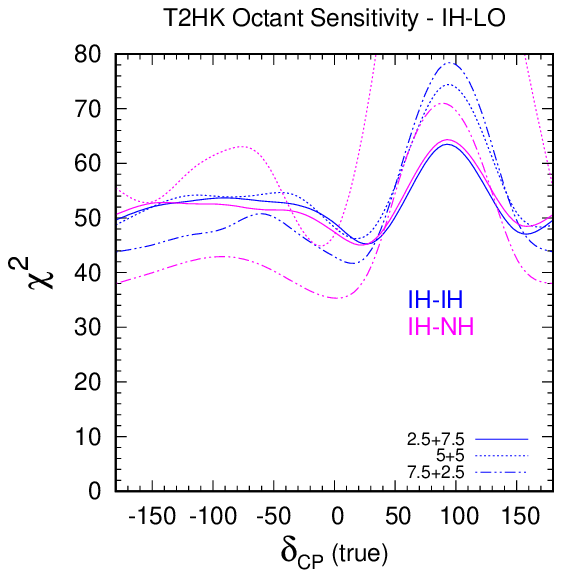}
                
                \hspace*{-1.0in}
                \includegraphics[width=0.5\textwidth]{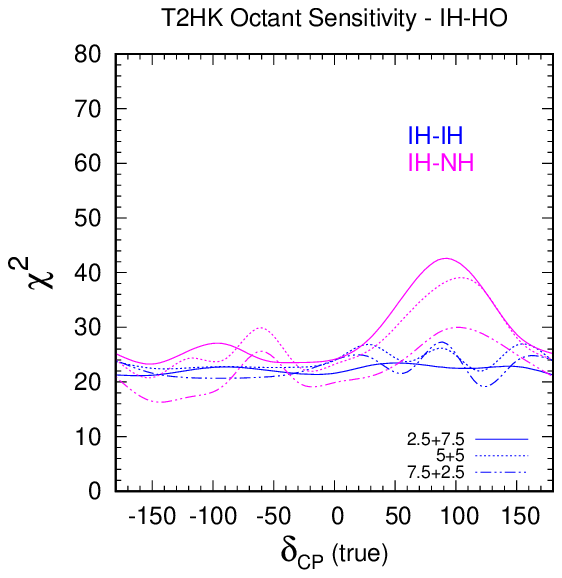}
        \end{tabular}
\caption{Octant sensitivity $\chi^2$ vs $\dcp$ plots for T2HK (true NH first row and true IH second row). The labels signify the $\nu$+$\bar{\nu}$ runs.}   
\label{oct-sens1}
\end{figure}

%fig. \ref{oct-sens2} shows the octant sensitivity of T2HKK for NH-LO, NH-HO,

Since the  octant degeneracies involving 
neutrinos and antineutrinos behave in an opposite way there are interesting 
interplay of run time ratios  for octant sensitivity 
This issue has been delved in detail in \cite{Nath:2015kjg} for 
DUNE. However for DUNE the WH solutions are largely absent and therefore 
one had to deal with only right hierarchy-wrong octant solutions. 
But for T2HK hierarchy degeneracy is also present and so one encounters 
wrong octant solutions for both right and wrong hierarchy, which further 
complicates the issue. 

\begin{figure}[h!]
%\vspace{-1.4cm} 
        \begin{tabular}{lr}
				 &	                 
                 \hspace*{-0.65in}
                \includegraphics[width=0.5\textwidth]{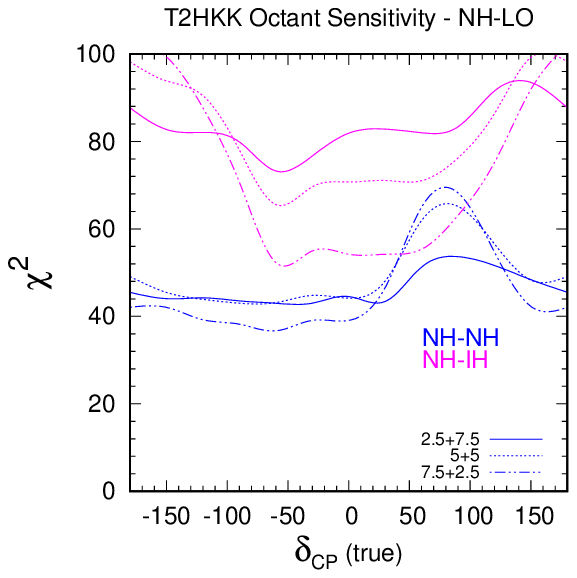}
                
                \hspace*{-1.0in}
                \includegraphics[width=0.5\textwidth]{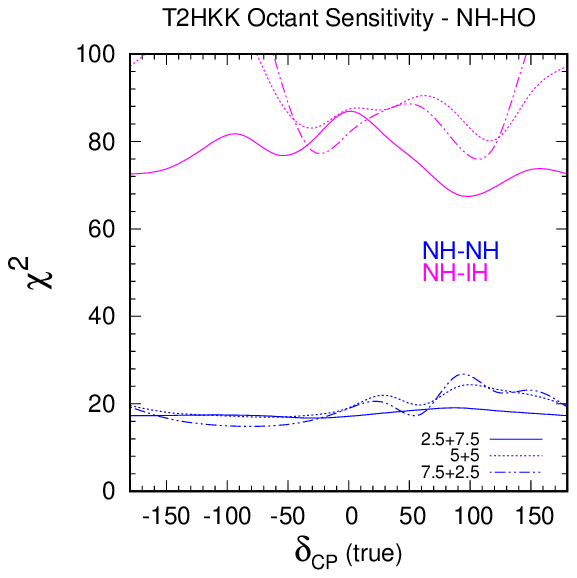}\\
                &
                \hspace*{-1.0in}
                \includegraphics[width=0.5\textwidth]{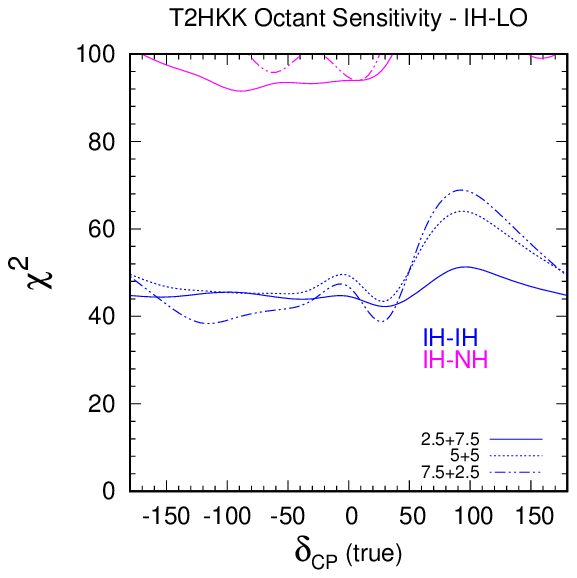}
                
                \hspace*{-1.0in}
                \includegraphics[width=0.5\textwidth]{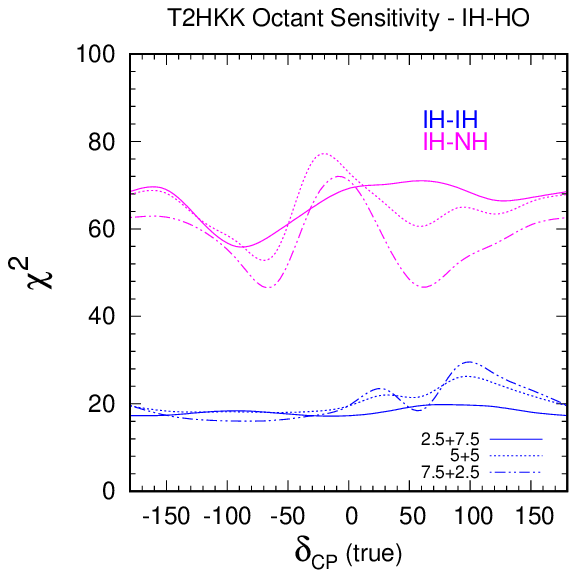}
        \end{tabular}
\caption{Octant sensitivity $\chi^2$ vs $\dcp$ plots for T2HKK (true NH first row and true IH second row). The labels signify the $\nu$+$\bar{\nu}$ runs.}   
\label{oct-sens2}
\end{figure}

Fig. \ref{oct-sens1} and fig. \ref{oct-sens2} show the octant 
sensitivity of T2HK and T2HKK experiments for various true values of $\delta_{CP}$ 
with respect to different true hierarchy--octant combinations. 
In each plot we show the octant sensitivity corresponding to three different 
run time ratios of 2.5$\nu$ + 7.5$\bar{\nu}$, 5$\nu$ + 5$\bar{\nu}$ and 
7.5$\nu$ + 2.5$\bar{\nu}$ shown by 
solid, dotted and dash-dotted curves respectively. 
The blue curves correspond to right hierarchy whereas magenta curves 
correspond to wrong hierarchy. 
Marginalization over hierarchy would chose the lower values of $\chi^2$ 
in each case. 

%Additionally, 
%in fig. \ref{oct-sens1} we also show the sensitivities of T2HK corresponding to the
%run time ratios of 0$\nu$ + 10$\bar{\nu}$(orange), 10$\nu$ + 0$\bar{\nu}$(brown) 
%to facilitate a better understanding of the physics behind the interplay of 
%neutrino and antineutrino runs to resolve the octant degeneracy.

In the top left panel of fig. \ref{oct-sens1} where true hierarchy--octant 
configuration is NH-LO ($\theta_{23}({true}) = 42^\circ$) 
it can be seen that the higher sensitivity for all 
values of $\delta_{CP}$ is obtained from the proposed run-time ratio of 
2.5$\nu$ + 7.5$\bar{\nu}$ for the right hierarchy in the LHP of $\dcp$ 
and over the whole range of $\dcp$ for the wrong hierarchy. 
If hierarchy is known to be NH then in the upper half plane 
5+5 or 7.5+2.5 gives better results. 
This can be understood
by relating to the corresponding probabilities shown in the top panel of 
fig. \ref{prob-plots}. 
In the LHP of the top right panel of fig. \ref{prob-plots}, we can see that 
the NH-LO(green) band for antineutrinos 
do not  suffer from  octant degeneracies  for both right and wrong hierarchy 
(the yellow and blue bands). Thus more antineutrinos help to have a higher 
octant sensitivity in both the cases.  
In the UHP for wrong hierarchy (the magenta curves) 
the  antineutrino run helps in removing the 
WH-WO solutions occurring with the same CP.
However since neutrinos  do not 
suffer from any degeneracy  between NH-LO and NH-HO in the UHP,
7.5+2.5 or 5+5 gives better 
results. 

For IH-LO, in the LHP neutrino has octant degeneracy for wrong as
well as right hierarchy  for wrong $\dcp$ 
as can be seen by comparing the red band with the  
yellow and the blue bands respectively in fig. \ref{prob-plots}. 
But antineutrino probabilities 
do not have any octant degeneracy. 
Thus the plan with more neutrinos is worse 
than that with equal or more antineutrinos. 
%For unknown hierarchy 5+5 fares better.
On the other hand in the UHP neutrino 
probability does not suffer from octant degeneracy as 
can be seen from the red band in the left panel 
of fig. \ref{prob-plots}  while the antineutrino 
probabilities have octant degeneracies. 
%for both correct 
%and incorrect octant.  
Thus for both right and wrong 
hierarchy the run with greater proportion of neutrinos 
is better. For wrong hierarchy 5+5 fares much better 
as in the LHP.

\begin{table}[h!]
%\begin{center}
%\vspace{-1cm}
\begin{tabular}{|c|c|c|c|c|c|c|}
\hline
 & $\theta_{23}(test)$(total minima) & $\theta_{23}(test)$(disappearance minima) & App($\nu$) & App($\bar{\nu}$) & Disapp($\nu$)+Dispp($\bar{\nu}$) & Total \\
\hline
10+0 & 43.5 & 43.3 & 13.94 & 0 & 0.6 & 14.54 \\
7.5+2.5 & 43.6 & 43.2 & 12.79 & 9.63 & 1.83 & 24.25 \\
5+5 & 43.7 & 43.2  & 11.44 & 13.66 & 3.08 & 28.17 \\
2.5+7.5 & 43.8 & 43.2 & 9.37 & 15.67 & 3.99 & 29.03  \\
0+10 & 44 & 43.1 & 0 & 15.79 & 5.17 & 20.96 \\
\hline
\end{tabular}
\caption{Contributions of $\chi^2$ from appearance and disappearance channels for true NH-HO 
and $\dcp = -90^\circ$.}
\vspace{10mm}
\label{tab:NH-HO}
%\end{center}
\end{table} 

For NH-HO in the LHP we see that the WH-WO solutions give a much 
higher $\chi^2$ and a run plan with more neutrinos perform better. 
This can be easily understood by comparing the yellow band and the 
red band from which it can be observed that there is no such 
degeneracy in the neutrino mode. 
However for the antineutrino probability
WH-WO-R$\dcp$ degeneracies can be observed.   
The NH-LO band (green) corresponding to the RH-WO 
solution  is closer to the NH-HO band and thus the 
$\chi^2$ for the RH case is lower. In this case also 
neutrino probabilities do not show any degeneracy. 
However, it is seen that even then the $2.5+7.5$ and $5+5$ 
give slightly better sensitivity even though antineutrinos 
have degeneracy for this case. 
In order to understand this  in table \ref{tab:NH-HO} we display the 
contribution of the different components to $\chi^2$. We ignore 
a constant prior term in this table. 
From the table we can see that for 10+0 i.e only neutrino run the 
contribution from the antineutrinos to the appearance channel $\chi^2$ 
is zero. As we decrease the neutrino component and increase the 
antineutrino component the appearance channel contribution from 
the neutrinos get reduced whereas the antineutrino contribution is
enhanced. Since neutrinos do not have any degeneracy for NH-HO 
whereas antineutrinos possess degeneracies with wrong CP the minima for 
neutrino and antineutrino do not come in the same position. 
The overall minima  is controlled by the neutrinos because of 
more statistics. But since this point is not the minima for 
the antineutrinos they give a large octant sensitive contribution. 
Thus the tension between neutrinos and antineutrinos help in raising the 
$\chi^2$ for the cases of mixed runs in spite of degeneracies in the 
antineutrino channel. There is another interesting feature which 
can be noticed in this table which is that the disappearance channels 
also contribute towards octant sensitivity. This is contrary to our 
expectations because the leading term in this channel goes as 
$1 - \sin^2\theta_{23} \sin^2\Delta m^2_{31}L/4E$ and does not have 
any octant sensitivity.

\begin{figure}[h!]          
\includegraphics[width=0.5\textwidth]{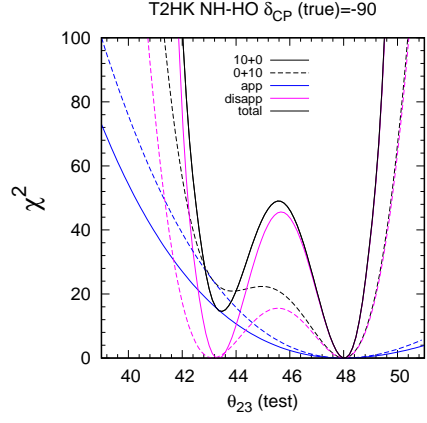}
\caption{Octant sensitivity $\chi^2$ vs test $\theta_{23}$ (degrees) plots for T2HK.}
\label{nu_anu_syn}
\end{figure}

To understand this behaviour in fig. \ref{nu_anu_syn} we plot the 
disappearance and appearance $\chi^2$ for neutrinos and 
antineutrinos separately as a function of test $\theta_{23}$. 
True $\theta_{23}$ is taken as $48^\circ$.  
It is seen from the figure and the table that the 
the global minima does not come at the disappearance minima but the 
appearance $\chi^2$ being a very steeply rising quantity tends to 
shift the global minima towards higher values of $\theta_{23}$.
This shift is more as the antineutrino component is increased 
since the antineutrino $\chi^2$ for appearance channel is steeper 
as compared to the neutrino $\chi^2$. 
Since the global minima is not the disappearance minima there 
is finite octant sensitive contribution from the disappearance 
channel as well. 

For IH-HO (blue band) in fig. \ref{prob-plots} there is WH-WO degeneracy with 
NH-LO (green band) at same CP value in the LHP for neutrinos. 
Antineutrinos on the other hand have degeneracy with both IH-LO (red band) 
and NH-LO (green band). However, still the runs with $2.5+7.5$ give 
similar results with $5+5$ and $7.5+2.5$. 
This can again be attributed to the  
tensions between neutrinos and antineutrinos as described for NH-HO. 
%{\bf although we have checked for +90}. 
For the UHP IH-HO has degeneracy (both WH-WO and RH-WO, green and red bands 
respectively) for neutrinos. On the other hand for antineutrinos there 
is no degeneracy.  Therefore $7.5+2.5$ give better sensitivities 
for the WH solutions. Note that for the WH solutions for neutrinos 
the degeneracy is at right CP value and hence neutrino and antineutrino 
minima both occur for right CP.
On the other hand 
for the right hierarchy solutions the neutrino minima occurs in the LHP 
while the antineutrino minima occurs in the UHP. 
This the tensions between neutrino and antineutrino occur here also
and all the three runtime proportions give similar results.

%Similar analysis and conclusions can be made for the other two configurations
%of NH-HO and IH-HO. In conclusion, one can observe that the octant sensitivity 
%of T2HK for any particular true hierarchy--octant configuration of 5$\nu$ + 5$\bar{\nu}$ 
%years run is more or almost equal to the sensitivity of the proposed runtime 
%of 2.5$\nu$ + 7.5$\bar{\nu}$ years.

Fig. \ref{oct-sens2} shows the octant sensitivity $\chi^2$ vs true $\dcp$ 
of T2HKK with NH-LO, NH-HO,
IH-LO, IH-HO as true hierarchy--octant configurations, where each plot shows 
right hierarchy (blue) and wrong hierarchy (magenta) curves for three different 
run time ratios 2.5$\nu$ + 7.5$\bar{\nu}$ (solid), 
5$\nu$ + 5$\bar{\nu}$ (dotted) and 7.5$\nu$ + 2.5$\bar{\nu}$ (dot-dashed). 
Note that we will get the solutions with lower $\chi^2$ if we marginalize over the hierarchy. 

Owing to its longer baseline of 
1100 km we have shown in fig. \ref{mass-hier1} that T2HKK has high hierarchy sensitivity 
when compared to its counter-proposal T2HK. Thus the wrong hierarchy-octant 
solutions do not occur here.
As a result it can be understood from fig. \ref{oct-sens2} 
that the wrong hierarchy solutions NH-IH (magenta) curves
have comparatively higher $\chi^2$ in all the four cases 
and they will get removed once the hierarchy is marginalized. 

In section \ref{prob-discuss} we have given a detailed account of 
how the octant sensitivity at 1100 km baseline is very low because of 
the degeneracies. 
However, since T2HKK is a hybrid setup with one detector placed at 295 km 
and another at 1100 km, the probabilities at both 
baselines govern the physics of this experiment. Thus, 
one can attribute the considerably large octant sensitivity arising in 
fig. \ref{oct-sens2} to be mainly coming from the 295 km baseline. Note 
that these can be only right hierarchy solutions as the wrong ones get
removed at 1100 km.

For instance, in the top left panel of the figure 
where we assume NH-LO as the true configuration,  
the proposed run time of 2.5$\nu$ + 7.5$\bar{\nu}$ (solid) 
or 5$\nu$ + 5$\bar{\nu}$ (dotted) 
give a better solution than 7.5$\nu$ + 2.5$\bar{\nu}$ 
in the LHP. This behaviour is the same for the right hierarchy 
solutions, shown by blue curves when true combination is NH-LO
as can be seen from the top left panel of fig. \ref{oct-sens1}. 
Similarly in the UHP 7.5$\nu$ + 2.5$\bar{\nu}$ or 5$\nu$ + 5$\bar{\nu}$
is better run, as neutrinos corresponding to NH-NH 
do not suffer from any degeneracy as seen from fig. \ref{prob-plots}.

To sum it up, for all true hierarchy-octant combinations,
the conclusions corresponding to NH-NH 
solutions (blue curves) of fig. \ref{oct-sens2} follow the physics 
at 295 km baseline i.e. T2HK and can be understood from the 
detailed description of the NH-NH solutions (blue curves) of 
fig. \ref{oct-sens1} presented before.

%It can be seen that for all the four combinations the sensitivity of 
%5$\nu$ + 5$\bar{\nu}$ run is more by $\chi^2 \approx (5-20)$ than that of 2.5$\nu$ + 7.5$\bar{\nu}$ 
%run for all true values of $\dcp$.
%
%\begin{figure}[h!]
%%\vspace{-1.4cm} 
%        \begin{tabular}{lr}
%				 &	                 
%                 \hspace*{-0.65in}
%                \includegraphics[width=0.5\textwidth]{t2hkk-octant-nhlo-all.eps}
%                
%                \hspace*{-1.0in}
%                \includegraphics[width=0.5\textwidth]{t2hkk-octant-nhho-all.eps}\\
%                &
%                \hspace*{-1.0in}
%                \includegraphics[width=0.5\textwidth]{t2hkk-octant-ihlo-all.eps}
%                
%                \hspace*{-1.0in}
%                \includegraphics[width=0.5\textwidth]{t2hkk-octant-ihho-all.eps}
%        \end{tabular}
%\caption{Octant sensitivity $\chi^2$ vs $\dcp$ plots for T2HKK (true NH first row and true IH second row). The labels signify $\nu$+$\bar{\nu}$ runs.}   
%\label{oct-sens2}
%\end{figure}

\subsection{Analysis of CP discovery potential}

\begin{figure}[h!]
%\vspace{-1.4cm} 
        \begin{tabular}{lr}
				 &	                 
                 \hspace*{-0.65in}
                \includegraphics[width=0.5\textwidth]{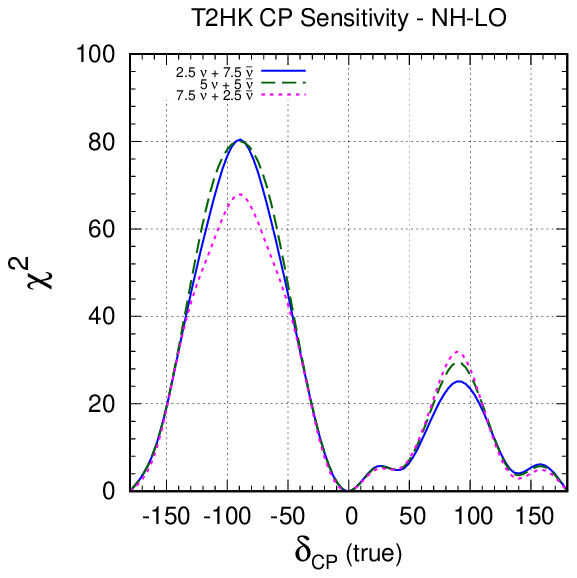}
                
                \hspace*{-1.0in}
                \includegraphics[width=0.5\textwidth]{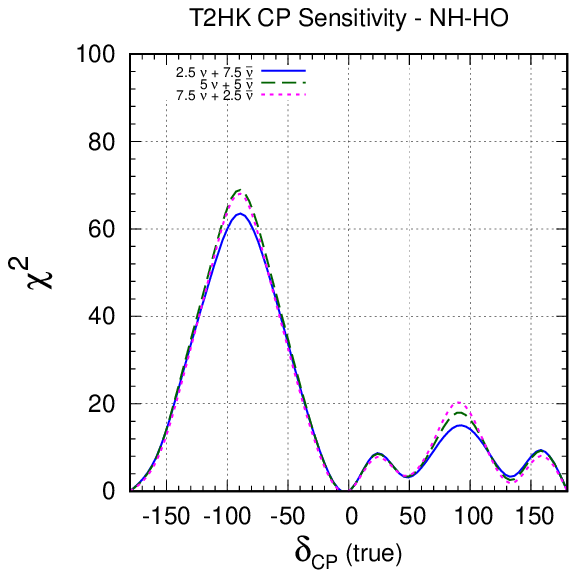}\\
                &
                \hspace*{-1.0in}
                \includegraphics[width=0.5\textwidth]{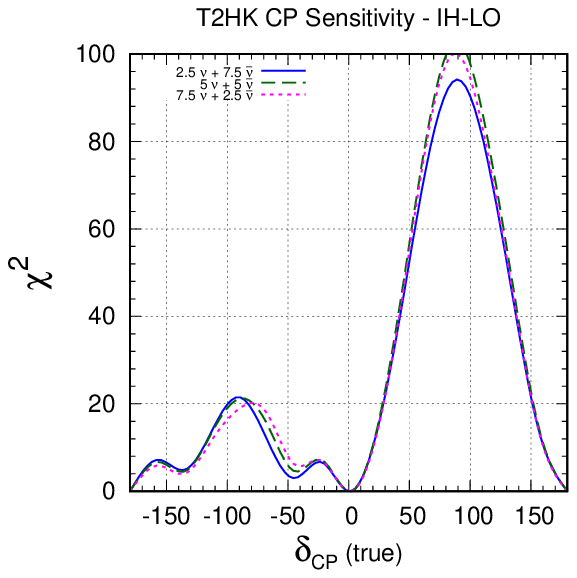}
                
                \hspace*{-1.0in}
                \includegraphics[width=0.5\textwidth]{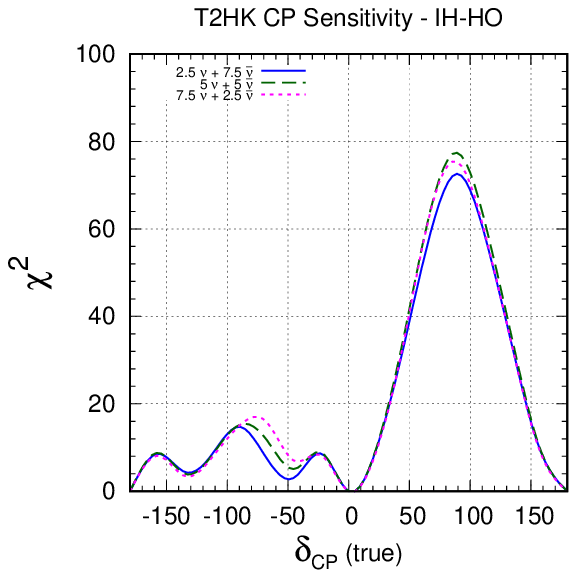}
        \end{tabular}
\caption{CP discovery potential $\chi^2$ vs $\dcp$ plots for T2HK (true NH first row and true IH second row). The labels signify the $\nu$+$\bar{\nu}$ runs.}   
\label{cp-sens1}
\end{figure}

\begin{figure}[h!]
%\vspace{-1.4cm} 
        \begin{tabular}{lr}
				 &	                 
                 \hspace*{-0.65in}
                \includegraphics[width=0.5\textwidth]{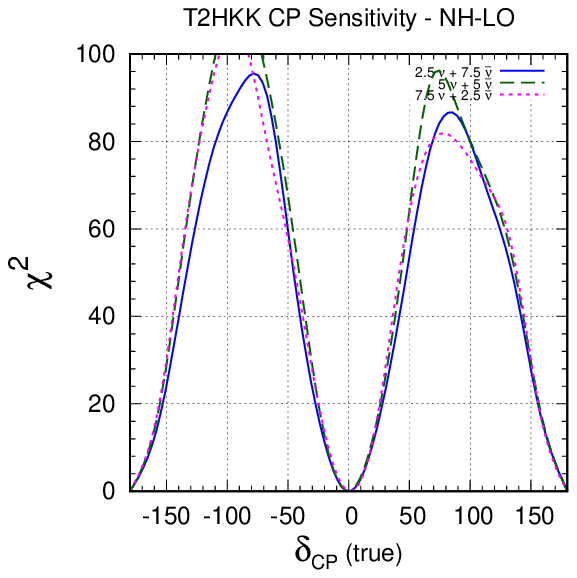}
                
                \hspace*{-1.0in}
                \includegraphics[width=0.5\textwidth]{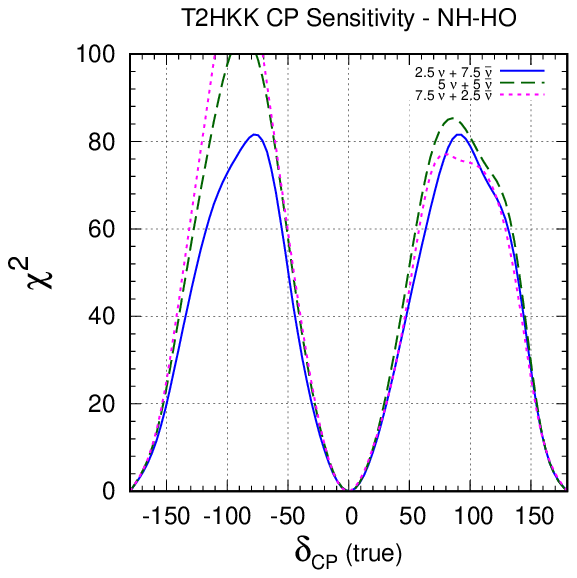}\\
                &
                \hspace*{-1.0in}
                \includegraphics[width=0.5\textwidth]{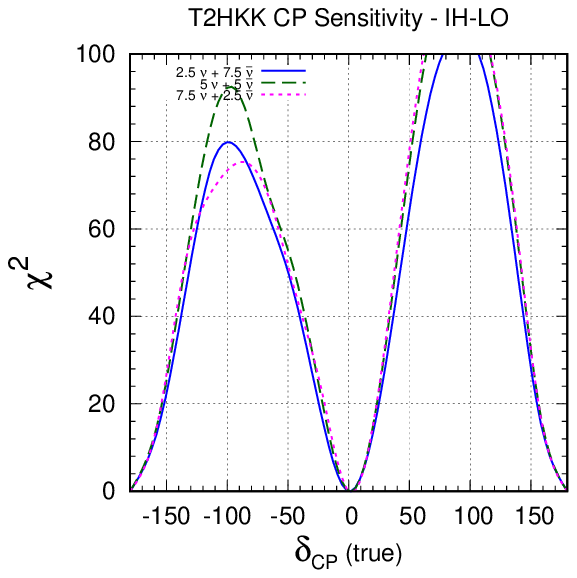}
                
                \hspace*{-1.0in}
                \includegraphics[width=0.5\textwidth]{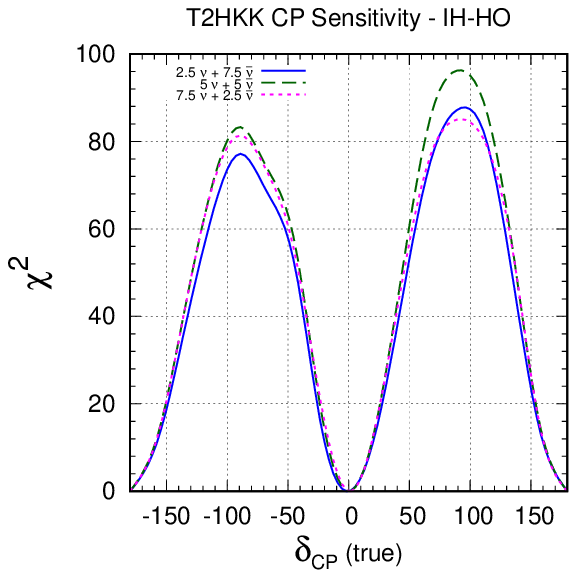}
        \end{tabular}
\caption{CP discovery potential $\chi^2$ vs $\dcp$ plots for T2HKK (true NH first row and true IH second row). The labels signify the $\nu$+$\bar{\nu}$ runs.}   
\label{cp-sens2}
\end{figure}

Fig. \ref{cp-sens1} shows the CP violation sensitivity of T2HK for 
different true hierarchy-octant combinations of NH-LO, NH-HO, IH-LO, IH-HO. 
Each plot shows the sensitivity corresponding to the run time ratios of 
2.5$\nu$ + 7.5$\bar{\nu}$ (blue-solid), 
5$\nu$ + 5$\bar{\nu}$ (magenta-dotted) and 7.5$\nu$ + 2.5$\bar{\nu}$ 
(green-dash-dotted). 
The figures show that in all cases the CP discovery potential is 
much less in one of the half-planes. This can be attributed to 
the presence of wrong hierarchy solutions as can be seen in fig 
\ref{mass-hier1}. 
However, we checked that the minima always occurs 
 with the right octant. It is known that the role of 
antineutrinos for T2HK baseline and energy is to remove the wrong 
octant solutions \cite{Ghosh:2015tan}. Therefore, more neutrinos 
will help because of enhanced statistics. This can be seen  from the 
plots in fig. \ref{cp-sens1} which show that the 1:1 or 3:1 ratios give 
slightly better results excepting true NH-LO. 
For this case, as can be observed from 
the left plot in the top panel,  around the maximal 
$\dcp$ in the LHP, the sensitivity of 5$\nu$ + 5$\bar{\nu}$ gives almost the same
result as the proposed 2.5$\nu$ + 7.5$\bar{\nu}$. However,
the sensitivity of 7.5$\nu$ + 2.5$\bar{\nu}$ i.e. for more neutrino run,
is lower. This is  because as seen 
from fig. \ref{prob-plots} NH-LO is degenerate with NH-HO at $\dcp = 0, \pm 180$
and IH-HO for right $\dcp$. But, in the case of antineutrinos NH-LO has degeneracy
only with IH-LO at right $\dcp$. As a result more antineutrinos are helping 
the sensitivity.

%\textbf{In the UHP we see the $\chi^2$ is lowered because of wrong hierarchy-right octant
% solutions.
%Here around maximal $\dcp$ it can be seen that, 
%more of neutrino run 7.5$\nu$ + 2.5$\bar{\nu}$ has 
%high sensitivity. The role of antineutrinos is to 
%remove octant degeneracy for T2HK baseline. But as in the UHP 
%the sensitivity is effected by the right octant solutions 
%the role played by more antineutrinos becomes redundant. Instead,
% the high statistics of neutrino runs increase the CP discovery 
% potential for this case.}
 
%\textbf{In the UHP the CP sensitivity is impaired due to 
%% WH solutions. 
%However, we checked that the minima occurs 
% with the right octant. It is known that the role of 
%antineutrinos for T2HK baseline and energy is to remove the wrong 
%octant solutions \cite{}. Therefore, more neutrinos 
%will help because of enhanced statistics.}

%Coming to the true NH-HO combination we can see that 
%7.5$\nu$ + 2.5$\bar{\nu}$ is doing better both in the LHP and the UHP. 
%This can be clearly seen from the yellow band of fig. \ref{prob-plots}
%for both neutrino and antineutrino probabilities. 

%Similar conclusions can be drawn for IH-LO, IH-HO. However for these cases
% UHP is effected by the wrong hierarchy-right octant solutions and hence the 
%sensitivity goes low. A general conclusion that a mixed 
%5$\nu$ + 5$\bar{\nu}$ run gives better or at times 
%comparable CPV sensitivity than the proposed 
%2.5$\nu$ + 7.5$\bar{\nu}$ run can be drawn.

Fig. \ref{cp-sens2} displays the CP discovery potential of T2HKK.
One major difference between the sensitivity of T2HK (in fig. \ref{cp-sens1}
and T2HKK is that the UHP of true NH and the LHP of the true IH no longer 
suffer from degeneracy coming from the wrong hierarchy solutions as these
get removed at the 1100 km baseline.  On the other hand the wrong octant 
solutions are addressed by the 295 km baseline as seen earlier. 
%From all the four plots one can conclude that the 
Overall 
CP discovery potential of 5$\nu$ + 5$\bar{\nu}$ is comparatively 
better than the proposed run of 2.5$\nu$ + 7.5$\bar{\nu}$ years 
in all the four cases. 

\section{Conclusions}

In this paper we have performed a comprehensive 
comparative analysis of  the hierarchy, octant 
and CP discovery sensitivities of the proposed high statistics experiments 
T2HK, T2HKK and DUNE. We have also obtained the sensitivities by including the 
projected full runs of T2K and \nova. 
We have discussed the  underlying 
physics issues which could cause the differences in the sensitivities.
In particular we present a detailed discussion on the octant sensitivities 
of the T2HKK experiment bringing out the salient features of the relevant probabilities  
near the second oscillation maxima.  

Our study shows that with their proposed fiducial volume and run time, 
T2HKK and DUNE experiments can achieve very high hierarchy sensitivity. 
The T2HK experiment on the other hand cannot resolve hierarchy-$\dcp$ 
degeneracy in the range $-180^\circ < \dcp <0$ for IH and 
$0<\dcp<180^\circ$ for NH. Hence in these unfavourable regions the hierarchy
sensitivity can reach maximum $3\sigma$.
In the favourable half-plane of $\dcp$ more than 
$5\sigma$ sensitivity is possible for T2HK stand-alone and more than 
$6\sigma$ including T2K and \nova information. 
Highest hierarchy sensitivity can be achieved by DUNE. 
On the other hand T2HK has the highest octant sensitivity among all the 
three experiments. T2HKK and DUNE have comparable octant sensitivity.
The CP discovery potential on the other hand is best for T2HKK. 
T2HK gives comparable sensitivity to T2HKK for favourable values of $\dcp$.
However for unfavourable $\dcp$ values due to wrong hierarchy-wrong CP solutions
the CP discovery potential suffers. DUNE does not have hierarchy degeneracy.
But it's sensitivity to CP discovery potential is lower because of 
lower volume as compared to T2HK and T2HKK. 

We also compute  the optimal exposure for $5\sigma$ hierarchy and octant 
sensitivity of all the experiments both stand-alone and in conjunction with 
T2K and \nova results.
% We present these results for the 
%most optimistic case. 
%For more favourable cases the optimum exposure needed
%will be even less. 
The sensitivity study is done for two representative
configurations assuming true NH-LO 
($\theta_{23}= 42^\circ$) and IH-HO ($\theta_{23}= 48^\circ$). 
For hierarchy sensitivity  better result is obtained for IH-HO.
We find that  for this case DUNE can attain $5\sigma$ sensitivity 
in approximately 3 years (equal neutrino and antineutrino run) and with T2K and 
\nova information it is just 1.5 years with a 40 kt volume. 
%For 10 kt volume, i.e the first phase of DUNE can achieve this in 
%8 years including the T2K and \nova information. 
T2HKK with it's proposed volume can achieve this in 3.6 and 2.3 years 
respectively.  
Optimum exposure required to obtain $5\sigma$ octant sensitivity 
is found to be less when we assumed true NH-LO.
In this case T2HK can 
deliver the result in 3.8 years and with T2K and \nova 
the same can be achieved within 2.7 years. 
On the other hand, T2HKK and DUNE would need more than 10 years of exposure to 
attain this goal. 
We also present the optimal volume for which $3\sigma$ CP violation
discovery  sensitivity 
can be reached for 60\% fraction of  $\dcp$ values for both NH and IH.
This is found to be 400 kt-yr for T2HKK for both true NH and IH. 
For DUNE the optimum exposure is 
500 kt-yr (400 kt-yr) for DUNE for true NH (IH). 
Since T2HKK has larger volume, it can therefore achieve the same goal 
in lesser time. 

Finally, we study if the 1:3 neutrino-antineutrino run as proposed 
by T2HK and T2HKK is the best option or the alternative ratios of
 3:1/1:1 give better sensitivities. 
We find that for hierarchy sensitivity of T2HK this does not make much 
difference. But for T2HKK for certain range of $\dcp$
values 3:1 ratio for neutrinos
and antineutrinos give better hierarchy sensitivity. Additionally,
we observed that 
%for most of the values of true $\dcp$ 
the $\nu$:$\bar{\nu}$ run time ratio of 1:1 results almost as good 
(or sometimes better)
sensitivity as the proposed 3:1 ratio.
For octant sensitivity there is an interesting interplay between the 
$\nu$,$\bar{\nu}$ run time ratios since the octant degeneracy is
different for neutrinos and antineutrinos. However, 
$(\nu+\bar{\nu})$ combination can raise the $\chi^2$ since their 
corresponding minimas do not occur at the same point. Apart from this the
tension between the appearance and the disappearance channel shifts the minima 
from the disappearance minima thus leading to an octant sensitive contribution 
from this channel.  
For T2HK experiment when 
true hierarchy is IH and test hierarchy is marginalized we find that
the octant sensitivity for all three run time ratios is similar. But when 
we assume true NH-LO the sensitivity of 3:1 is poorer than 1:3 and 1:1 which 
are almost similar.  But for NH-HO in the UHP 1:1 and 3:1 ratios  perform 
better. 
For T2HKK, overall there are parameter spaces near $\dcp = +90^\circ$,
where 3:1 ratio give better results. Otherwise all the three scenarios are 
in the same ballpark.
The CP discovery potential for T2HK is almost the same for the  three 
run time ratios. For T2HKK the 3:1 and 1:1 ratios give slightly better results.
 
To conclude, all the three proposed next generation experiments
DUNE, T2HK and T2HKK hold the promise to measure the unknown 
neutrino oscillation parameters with a high sensitivity.

%From the detailed analysis of our results we can conclude that DUNE, T2HK \& T2HKK experiments are very high precision experiments and are expected to be very successful in the determination of the unknowns of neutrino oscillation regime. If the data from the ongoing experiments like T2K and NO$\nu$A are combined with the future experiments then the fiducial volumes of the detectors required for the experiments can be reduced hence can reduce the cost of the future experiments. 
%We also conclude that the best configuration for the T2HK \& T2HKK experiment is if we have one detector at 295km and the other at 1100km $1.5^\circ$ off-axis, although the proposed runtime for neutrino and antineutrino mode for the experiment is 2.5$\nu$ + 7.5$\bar{\nu}$ but from our analysis we observe that 7.5$\nu$ + 2.5$\bar{\nu}$ run time configuration is better than the proposed runtime configuration. Since DUNE is also a proposed experiment so on adding DUNE with T2HKK run with runtime configuration of 7.5$\nu$ + 2.5$\bar{\nu}$ (fig. \ref{t2hkk+dune}) will be able to solve all the three unknowns of neutrino oscillation physics with high precision. We see that adding DUNE with T2HKK improves the statistics and can resolve the neutrino oscillations regime at $5\sigma$ confidence level. 
\section{Acknowledgement} Authors would like to thank Monojit Ghosh 
for many helpful discussions.

\bibliographystyle{unsrt}
\bibliography{neutrinoref.bib}
 
\end{document}